\documentclass[preprint,amsmath,amssymb,prb,longbibliography]{revtex4-1}

\usepackage{amssymb}
\usepackage{amsfonts}
\usepackage{color}
\usepackage{dsfont}
\usepackage[normalem]{ulem}
\usepackage{dcolumn}
\usepackage{mathtools}
\usepackage{siunitx}
\usepackage[colorlinks=true,linkcolor=blue,citecolor=blue,urlcolor=blue]{hyperref}
\usepackage{graphicx}
\usepackage{blindtext}
\usepackage{outlines}
\usepackage{enumitem}
\usepackage{soul}
\setlist[enumerate,2]{label=\roman*)}
\setlist[enumerate,3]{label=\alph*)}
\usepackage{multirow}

\newcommand{\dd}{\ensuremath{\text{d}}}

\newcommand{\iu}{\mathrm{i}}

\newcolumntype{d}[1]{D{.}{.}{#1}}
\definecolor{orange}{rgb}{1,0.5,0}
\definecolor{darkgreen}{RGB}{0,100,0}

\makeatother

\begin{document}

\title{Spin dynamics of $3d$ and $4d$ impurities embedded in prototypical topological insulators}
\author{Juba Bouaziz}
\email{j.bouaziz@fz-juelich.de}
\author{Manuel dos Santos Dias}
\author{Filipe Souza Mendes Guimar\~aes}
\author{Samir Lounis}
\affiliation{Peter Gr\"unberg Institut and Institute for Advanced Simulation, 
Forschungszentrum  J\"ulich and JARA, 52425 J\"ulich, Germany}

\date{\today}

\begin{abstract}
Topological insulators are insulating bulk materials hosting conducting surface states. 
Their magnetic doping breaks time-reversal symmetry and generates numerous interesting 
effects such as dissipationless transport. Nonetheless, their dynamical properties are still
poorly understood. Here, we perform a systematic investigation of transverse 
spin excitations of $3d$ and $4d$ single impurities embedded in two prototypical topological 
insulators (Bi$_2$Te$_3$ and Bi$_2$Se$_3$). The impurity-induced states within the bulk gap
of the topological insulators are found to have a drastic impact on the spin excitation
spectra, resulting in very high lifetimes reaching up to \textit{microseconds}. An intuitive
picture of the spin dynamics is obtained by mapping onto a generalized Landau-Lifshitz-Gilbert
phenomenological model. The first quantity extracted from this mapping procedure is the magnetic
anisotropy energy, which is then compared to the one provided by the magnetic force theorem.
This uncovers some difficulties encountered with the latter, which can provide erroneous results
for impurities with a high density of states at the Fermi energy. Moreover, the Gilbert damping
and nutation tensors are obtained. The nutation effects can lead to a non-negligible shift 
in the spin excitation resonance in the high-frequency regime. 
Finally, we study the impact of the surface state on the spin dynamics, which may be severely 
altered due to the repositioning of the impurity-induced state in comparison to the bulk case.
Our systematic investigation of this series of magnetic impurities sheds
light on their spin dynamics within topological insulators, with implications for available
and future experimental studies as, for instance, on the viability of using such impurities for solid-state qubits. 
\end{abstract}

\maketitle

\section{Introduction}

The ever-increasing need for higher storage density oriented research towards 
the miniaturization of magnetic memories, constricted by the super-paramagnetic 
limit~\cite{Shiroishi:2009}. The realization of smaller magnetic bits requires 
materials with a high magnetic anisotropy energy (MAE),  originating from 
the relativistic spin-orbit interaction. The extreme limit for high-density 
magnetic storage consists of a single atomic bit~\cite{Natterer:2017}, for which 
quantum effects can be predominant.
Therefore, a deep fundamental understanding underlying the stability mechanisms
is crucial for future technological applications. Moreover, the manipulation of 
these magnetic units relies on external time-dependent fields, with their dynamical 
properties being of prime relevance as well. 

The standard tool for probing the dynamical magnetic properties (\textit{i.e.} spin 
excitations) of single atoms is the inelastic scanning tunneling spectroscopy 
(ISTS). It was employed to investigate magnetic adatoms on non-magnetic surfaces~\cite{Heinrich:2004,Bryant:2013,Oberg:2013,Rossier:2009,Loth:2010,Balashov:2009,Khajetoorians:2011,Chilian:2011,Khajetoorians:2013,Donati:2013}. The spin excitations signature in the differential conductance 
($\frac{\dd I}{\dd V}$, with $I$ being the tunneling current and $V$ the applied voltage) consists of 
step-like features at the excitation frequencies. They are determined by the 
applied external magnetic field and the MAE, which can also be accessed via other 
experimental methods such as X-ray magnetic circular dichroism (XMCD)~\cite{Honolka:2012,Gambardella:2003}. 
The nature of both the substrate and the adsorbate play a major role in the determination
of the resonance frequency and lifetime of the excitation. 

Several theoretical investigations of spin excitations of magnetic atoms deposited 
on nonmagnetic surfaces have been performed. In the limit of weak coupling (\textit{i.e.} 
low hybridization) between the adsorbate and the substrate, the ISTS spectra can 
be interpreted employing a Heisenberg model with localized atomic moments possessing 
an integer (or half integer) spin. Such a scenario occurs when the substrate is of 
insulating or semi-conducting nature~\cite{Rossier:2009,Fransson:2009,Fransson:2010}. 
When the coupling to the substrate is strong, the hybridization effects must be taken 
into account and a more accurate description of the electronic structure is required. 
This was achieved 
using real-space first-principles calculations in the framework of the Korringa-Kohn-Rostoker 
Green function (KKR-GF) method, which was extended to the dynamical regime~\cite{Samir:2010,Samir:2011,Samir:2013,Manuel:2015} relying on 
time-dependent density functional theory (TD-DFT) in its linear response formulation~\cite{Gross:1985}. 

Topological insulators are intermediate between metallic and insulating substrates, 
consisting of bulk insulators hosting conducting topologically protected
surface states~\cite{Hasan:2010,Qi:2011,Zhang:2009}. The magnetic doping of topological insulators 
breaks time-reversal symmetry and generates exotic phenomena such as the quantum anomalous 
Hall effect~\cite{Chao:2016,Islam:2018}. In this case, one also expects a rather low but finite 
hybridization (with the surface state) in the region of the bulk 
gap, leading to unconventional dynamical behaviour. For instance, the magnetization 
dynamics of a ferromagnet coupled to the surface state of a three-dimensional (3D) topological 
insulator has already been investigated, and an anomalous behaviour 
in the ferromagnetic resonance was predicted~\cite{Yokoyama:2010}. Other studies with a similar focus 
were done in Refs.~\onlinecite{Tserkovnyak:2012,Garate:2010,Ueda:2012,Dora:2015}.
Furthermore, arrays of magnetic adatoms interacting with a topological surface state were  
considered in Ref.~\onlinecite{Chotorlishvili:2014}, with the surface magnons following a
linear dispersion, very unusual for a ferromagnetic ground state. Moreover, the electron spin resonance
of single Gd ions embedded in Bi$_2$Se$_3$ was examined in Ref.~\onlinecite{Garitezi:2015}. 
The temperature dependence of the g-factor was investigated and the coexistence of a 
metallic and an insulating phase (dual character) was reported.

In this paper, we systematically investigate the spin dynamics of $3d$ and $4d$ 
single impurities embedded in prototypical 3D topological insulators, namely 
Bi$_2$Te$_3$ and Bi$_2$Se$_3$. Thin film (with a topological surface state) and 
inversion symmetric bulk (insulating) geometries are considered. For an accurate 
description of the dynamical electronic properties of these impurities,
we employ linear response TD-DFT as implemented in the KKR-GF method~\cite{Samir:2010,Samir:2011,Manuel:2015}. 
We compute the dynamical transverse magnetic susceptibility, which represents the 
magnetic response of the system to frequency-dependent transverse magnetic fields. 
It incorporates the density of spin excitations and can be connected to  
ISTS measurements~\cite{Schweflinghaus:2014}. The spin excitation spectra we obtain reveals 
astonishing results, with lifetimes spanning six orders of magnitude: from \textit{picoseconds}
to \textit{microseconds} for Fe and Mn impurities embedded in Bi$_2$Se$_3$, respectively. 
These contrasting values of the lifetimes 
correlate with the presence (or absence) of in-gap states in the impurity 
local density of states (LDOS) near the Fermi energy~\cite{Juba:2018}. Next we gain further insight
on the magnetization dynamics by mapping the transverse dynamical magnetic 
susceptibility to the phenomenological Landau-Lifshitz-Gilbert (LLG) 
equation~\cite{Gilbert:2004}. A generalized formulation of the LLG equation 
including tensorial Gilbert damping $\mathcal{G}$ and nutation $\mathcal{I}$ is employed~\cite{Bhattacharjee:2012}. The 
static limit of the response function via the LLG formulation was used 
to extract the MAE. The latter is then compared to the values obtained 
with conventional ground state methods relying on the magnetic force theorem: band energy differences~\cite{Oswald:1985,LKAG:1987,Daalderop:1990} 
and torque method~\cite{Wang:1996}. A connection between the MAE 
obtained within the linear response theory and the torque method using
small deviations is established. Moreover, for elements with high resonance 
frequencies, the signature of the nutation is observed as a resonance shift,
proving that inertial effects are relevant at such high precession rates~\cite{Sack:1957,Ciornei:2011,Bhattacharjee:2012}. 
Finally, we compare the LLG parameters obtained when the $3d$ and $4d$ impurities are embedded 
in the bulk and at the surface of Bi$_2$Te$_3$.
Our results show that the modification of the in-gap state due to the presence of the surface state 
may play a major role in the dynamics depending on the nature of the impurity.

This paper is structured as follows. Sec.~\ref{theory_section_tddft_llg} is 
dedicated to the description of the linear response TD-DFT approach employed
to compute the spin excitation spectra. It also includes the mapping of the 
transverse dynamical magnetic susceptibility into the generalized phenomenological LLG model 
and the different methods used to compute the MAE. Sec.~\ref{elec_str_3d_4d}
is devoted to the analysis of the electronic structure and the ground state 
properties of $3d$ and $4d$ transition metal impurities embedded in Bi$_2$Te$_3$ 
and Bi$_2$Se$_3$. In Sec.~\ref{Mca_3d_4d_Bi2Te3}, we present the MAE for the considered 
magnetic impurities and explain the discrepancies between the different methods. 
Sec.~\ref{SE_3d_4d_allhosts} contains a detailed discussion of the 
spin excitation spectra of $3d$ and $4d$ impurities embedded at the surface of 
both Bi$_2$Te$_3$ and Bi$_2$Se$_3$. The fitted LLG parameters are given as well, 
which are interpreted in terms of the impurity LDOS. Finally, in 
Sec.~\ref{spin_dyn_bulk_VS_surface}, the dynamical properties of the $3d$ 
impurities in the bulk and at the surface are compared. The contribution 
of the topological surface state for each impurity is then analyzed. 

\section{Theoretical description}
\label{theory_section_tddft_llg}
The description of the spin excitations of the investigated systems relies on linear response 
TD-DFT~\cite{Gross:1985,Samir:2010,Samir:2015,Manuel:2015}. 
The central quantity in our approach is the dynamical magnetic susceptibility, 
which displays poles at the excitation energies of the system. The calculations 
are performed in two steps: First we determine the ground state of the system using 
conventional DFT calculations; then, we compute the dynamical 
response of the system to an external perturbing time-dependent magnetic 
field.
To gain further physical insights into the results, we also describe how to map the results
of TD-DFT calculations onto an extended phenomenological LLG model.
Lastly, we compare the MAE obtained from the dynamical calculations with the ones computed
from DFT calculations in different ways.

\subsection{Density functional theory}
The ground state DFT simulations are done using the 
KKR-GF method~\cite{Papanikolaou:2002,Bauer:2014}
in the atomic sphere approximation (ASA) including the full charge density, 
and the exchange-correlation potential is taken in the local spin density 
approximation (LSDA)~\cite{Vosko:1980}. The spin-orbit interaction is included 
in a self-consistent fashion within the scalar relativistic approximation. 
Since we investigate impurities embedded in periodic crystals, we perform 
two types of calculations. The ground state of the clean host is determined 
first. Then, the impurities are self-consistently embedded in its crystalline 
structure. The host crystals investigated in this work consist of Bi$_2$Te$_3$ 
and Bi$_2$Se$_3$. The bulk unit cell contains five atoms (one quintuple layer) 
in a rhombohedral structure (space group {R}$\bar{3}$m)~\cite{Wei:2010}. 
The corresponding self-consistent calculations employ a  $30 \times 30 \times 
30$ $k$-mesh. The surface is simulated using a slab containing six quintuple 
layers and $60 \times 60$ $k$-points, as in our previous work~\cite{Juba:2018}. 

\subsection{Time-dependent density functional theory}
The dynamical magnetic susceptibility encodes the spin excitation 
spectra. It describes the linear change in the spin magnetization density $\delta\vec{M}(\vec{r},\omega)$ upon
the application of a frequency-dependent external magnetic field 
$\delta \vec{B}(\vec{r},\omega)$ as
\begin{equation}
\delta M_{\alpha}(\vec{r},\omega) = \sum_{\gamma}\int\!\dd\vec{r}^{\,\prime}\,
\chi_{\alpha\gamma}(\vec{r},\vec{r}^{\,\prime},\omega)\,\delta 
B_{\gamma}(\vec{r}^{\,\prime},\omega)\quad,
\label{liner_response}
\end{equation}
where $\alpha,\gamma \in \{x, y, z\}$. For a specific direction of $\vec{M}(\vec{r})$, the susceptibility 
tensor can be divided into longitudinal and transversal blocks.
In presence of the spin-orbit interaction or magnetic non-collinearity, the two blocks are coupled. 
However, for the systems that we analyze in this paper, the coupling is negligible and we focus only  
on the transversal magnetic response of systems (the $xy$ block when the magnetic moment 
is along the $z$-direction). Within TD-DFT, the magnetic susceptibility
$\chi_{\alpha\beta}(\vec{r},\vec{r}^{\,\prime},\omega)$ is determined starting from 
the non-interacting magnetic susceptibility of the Kohn-Sham system, 
${\chi}_{\alpha\beta}^\text{KS}(\vec{r},\vec{r}^{\,\prime},\omega)$, using a Dyson-like 
equation~\cite{Gross:1985,Samir:2010,Manuel:2015}:
\begin{equation}
\begin{split}
\chi_{\alpha\beta}&(\vec{r},\vec{r}^{\,\prime},\omega) = 
\chi^\text{KS}_{\alpha\beta}(\vec{r},\vec{r}^{\,\prime},\omega) \, + \\
& \sum_{\gamma\mu=x,y} \int\!\dd\vec{r}_{1}\,\dd\vec{r}_{2}\,
\chi^\text{KS}_{\alpha\gamma}(\vec{r},\vec{r}_{1},\omega)\,
K_{\gamma\mu}^\text{xc}(\vec{r}_{1},\vec{r}_{2},\omega) 
\,\chi_{\mu\beta}(\vec{r}_2,\vec{r}^{\,\prime},\omega)
\hspace{2mm},
\end{split}
\label{master_tddft}
\end{equation}
where $\alpha,\beta,\gamma,\mu \in \{x,y\}$ and $K_{\gamma\mu}^\text{xc}(\vec{r},\vec{r}^{\,\prime},\omega)$ 
is the transverse part of the exchange-correlation kernel, with $K^\text{xc}_{\gamma\mu}(\vec{r},\vec{r}^{\,\prime},\omega)= \delta_{\gamma\mu} K^\text{xc}_\perp(\vec{r},\vec{r}^{\,\prime},\omega)$. In the framework of the adiabatic 
LDA~\cite{Gross:1985,Liu:1989}, $K_\perp^\text{xc}(\vec{r},\vec{r}^{\,\prime},\omega) = \delta(\vec{r}-\vec{r}^{\,\prime})\,2B_{\text{xc}}(\vec{r}\,)/M(\vec{r}\,)$ is 
frequency-independent and local in space. The dynamical Kohn-Sham susceptibility is evaluated from the single particle Green function $\boldsymbol{G}(\vec{r},\vec{r}^{\,\prime},\varepsilon)$ (defined in Eq.~\eqref{def_GF}) as: 
\begin{equation}
\begin{split}
\chi^\text{KS}_{\alpha\beta}(\vec{r},\vec{r}^{\,\prime},\omega) & = -\frac{1}{\pi}\int_{-\infty}^{\varepsilon_\text{F}}
\dd\varepsilon\,\text{Tr}\,{\{}\boldsymbol{\sigma}_{\alpha}\,
\boldsymbol{G}(\vec{r},\vec{r}^{\,\prime},\varepsilon+\omega+\iu 0)\,\boldsymbol{\sigma}_{\beta}
\,\text{Im}\,\boldsymbol{G}(\vec{r}^{\,\prime},\vec{r},\varepsilon)\\ & +
\boldsymbol{\sigma}_{\alpha}
\,\text{Im}\,\boldsymbol{G}(\vec{r},\vec{r}^{\,\prime},\varepsilon)\,\boldsymbol{\sigma}_{\beta}
\,\boldsymbol{G}(\vec{r}^{\,\prime},\vec{r},\varepsilon-\omega-\iu 0)\}\quad.
\end{split}
\end{equation}
Since the frequency range of interest is relatively 
low~\cite{Samir:2015,Manuel:2015}, the frequency dependence of the Kohn-Sham susceptibility is incorporated via a Taylor expansion as
\begin{equation}
\chi^\text{KS}_{\alpha\beta}(\vec{r},\vec{r}^{\,\prime},\omega) \approx \chi^\text{KS}_{\alpha\beta}(\vec{r},\vec{r}^{\,\prime},0)
 + \left.\omega\,\frac{\dd\chi^{\text{KS}}_{\alpha\beta}(\vec{r},\vec{r}^{\,\prime},\omega)}{\dd\omega}\right|_{\omega=0}
 + \left.\frac{\omega^{2}}{2}\,\frac{\dd^2\chi^{\text{KS}}_{\alpha\beta}(\vec{r},\vec{r}^{\,\prime},\omega)}{\dd\omega^{2}}\right|_{\omega=0}\quad.
\label{Taylor_exp_KSsusc}
\end{equation}
$\chi^\text{KS}_{\alpha\beta}(\vec{r},\vec{r}^{\,\prime},0)$ being the static Kohn-Sham susceptibility.
Moreover, for a system with uniaxial 
symmetry, the transversal excitations can be summarized in the spin-flip magnetic susceptibility~\cite{Manuel:2015}
\begin{equation}
\chi_{+-}(\vec{r},\vec{r}^{\,\prime},\omega) = \frac{1}{4}\left[\chi_{xx}(\vec{r},\vec{r}^{\,\prime},\omega) 
+ \iu\chi_{xy}(\vec{r},\vec{r}^{\,\prime},\omega) - \iu\chi_{yx}(\vec{r},\vec{r}^{\,\prime},\omega) + 
\chi_{yy}(\vec{r},\vec{r}^{\,\prime},\omega) \right]\quad.
\label{chi_pm_susc}
\end{equation}
Further details on the computation of the Kohn-Sham susceptibility and exchange-correlation 
kernel can be found in Refs.~\onlinecite{Samir:2010,Samir:2015,Manuel:2015}. Finally, we 
can obtain an intuitive picture of the spin excitations via the spatial average of 
$\chi_{+-}(\vec{r},\vec{r}^{\,\prime},\omega)$ over a suitably-defined volume enclosing 
the magnetic impurity,
\begin{equation}
\chi_{+-}(\omega) = \int_{V}\!\!\dd\vec{r}\int_{V}\!\!\dd\vec{r}^{\,\prime}\,
\chi_{+-}(\vec{r},\vec{r}^{\,\prime},\omega)\quad,
\end{equation}
which corresponds to its net response to a uniform external magnetic field~\cite{Manuel:2015}.

\subsection{Generalized Landau-Lifshitz-Gilbert equation}
\label{LLG_section}
In order to develop a more intuitive picture of the magnetization dynamics, we make a
connection with a phenomenological model for the magnetization dynamics. We consider a 
generalized formulation of the Landau-Lifshitz-Gilbert 
(LLG) equation~\cite{Gilbert:2004} including a tensorial Gilbert damping $\underline{\mathcal{G}}$, 
as well as a nutation tensor $\underline{\mathcal{I}}$ accounting for inertial 
effects~\cite{Bhattacharjee:2012,Thonig:2012,Thonig:2017,Mondal:2017}. The latter
can be important at relatively high frequencies~\cite{Sack:1957,Ciornei:2011,Bhattacharjee:2012}. 
The equation of motion of the magnetic moment $\vec{M}(t) = 
\int_{V}\dd \vec{r}\,\,\vec{M}(\vec{r},t)$ then reads 
\begin{equation}
\frac{\dd\vec{M}}{\dd t} = -\gamma\,\vec{M}\times\left(\vec{B}^\text{eff} + 
\underline{\mathcal{G}}\cdot\frac{\dd\vec{M}}{\dd t} 
+ \underline{\mathcal{I}}\cdot\frac{\dd^{2}\vec{M}}{\dd t^{2}}\right)\quad.
\label{llg_genralized}
\end{equation}
Here $\gamma$ is the gyromagnetic ratio ($\gamma = 2$ in atomic units) and $\vec{B}^\text{eff}$ 
is the effective magnetic field acting on the magnetic moment. $\vec{B}^\text{eff}$
can be split into two contributions: $\vec{B}^\text{eff} = \vec{B}^\text{ext} + 
\vec{B}^\text{a}$, with $\vec{B}^\text{ext}$ being the external magnetic 
field, and $\vec{B}^\text{a}$ is an intrinsic anisotropy field which arises due 
to the spin-orbit interaction~\cite{Manuel:2015}. The relation between 
$\vec{B}^\text{a}$ and the magnetocrystalline anisotropy energy (MAE) $\mathcal{K}$ 
is detailed in Appendix~\ref{Append_A}. 

To establish a connection between the LLG equation and the transverse magnetic 
susceptibility computed using Eq.~\eqref{master_tddft}, we first consider that the local 
equilibrium direction is along the $z$-axis and apply a small time-dependent transverse 
magnetic field:
\begin{equation}
\vec{B}^\text{ext}(t) = \delta B_{x}(t)\,\vec{e}_{x} + \delta B_{y}(t)\,\vec{e}_{y}\quad;
\quad \text{with $\delta B_{x}(t),\delta B_{y}(t) \ll |\vec{B}^\text{a}|$}\quad.
\label{llg_external_field}
\end{equation}
Then, we linearize Eq.~\eqref{llg_genralized} with respect to transverse 
components of $\vec{B}^\text{ext}(t)$ and $\vec{M}(t)$, which becomes, in the 
frequency domain,
\begin{equation}
\sum_{\beta=x,y} \left(\frac{B^\text{a}_{z}}{M}\,\delta_{\alpha\beta}+\frac{\iu\omega}
{\gamma M}\,\epsilon_{\alpha\beta} + \iu\omega\,\mathcal{G}_{\alpha\beta} 
+ \omega^{2}\,\mathcal{I}_{\alpha\beta} \right)\,\delta M_{\beta}(\omega) = \delta B_{\alpha}
(\omega)\quad,
\label{LLG_generalized_1_adatom}
\end{equation}
with $\epsilon_{\alpha\beta}$ being the 2-dimensional Levi-Civita symbol 
($\epsilon_{xy} = +1$) and $\delta M_{\beta}(\omega)$ the  
$\beta$ component of the frequency dependent magnetization $\vec{M}(\omega)$. 
The preceding equation combined with Eq.~\eqref{liner_response} 
provides a direct connection between $\chi_{\alpha\beta}(\omega)$ obtained within 
TD-DFT and the phenomenological LLG parameters:
\begin{equation}
\begin{cases}
\left(\chi_{xx}(\omega)\right)^{-1} = -\frac{2\mathcal{K}_\text{Susc}}{M^{2}} 
- \frac{\iu\omega}{\gamma M} \mathcal{G}^{s}_{\parallel} - \frac{\omega^{2}}{\gamma M} 
\,\mathcal{I}^{s}_{\parallel}\quad,\\ 
\left(\chi_{xy}(\omega)\right)^{-1} 
= \frac{\iu\omega}{\gamma M} (1+\mathcal{G}^{a}_{\parallel}) + \frac{\omega^{2}}{\gamma M} 
\,\mathcal{I}^{a}_{\parallel}\quad,
\end{cases}
\label{fiting_inv_susc}
\end{equation} 
where $\mathcal{K}_\text{Susc}$ is the MAE, and the subscript indicates that this quantity 
is extracted from the static magnetic susceptibility obtained from the TD-DFT calculations.
$\mathcal{G}^{s}_{\parallel}$ $(\mathcal{I}^{s}_{\parallel})$ and 
$\mathcal{G}^{a}_{\parallel}$ $(\mathcal{I}^{a}_{\parallel})$ are the symmetric 
and anti-symmetric components of the Gilbert damping (nutation) tensor, 
respectively. A more detailed description of the Gilbert damping and 
nutation tensors for the uniaxial symmetry that applies to the systems under consideration
is provided in Appendix~\ref{Append_A}. The previous 
equation shows in a clear fashion that the static limit of $\chi_{xx}(\omega)$
is inversely proportional to the anisotropy. In the limit of small nutation,
the MAE is connected to the resonance frequency $\omega^\text{LLG}_\text{res}$ 
via (see Appendix~\ref{Append_A}) 
\begin{equation}
\omega^\text{LLG}_\text{res} = -\frac{\gamma}{\sqrt{1 + \big(\mathcal{G}^{s}_{\parallel}\big)^{2} 
+ 2\mathcal{G}^{a}_{\parallel}+ \big(\mathcal{G}^{a}_{\parallel}\big)^{2}}} 
\frac{2\mathcal{K}_\text{Susc}}{M_\text{s}}\quad.
\label{res_freq_llg}
\end{equation}
This is the resonance frequency for precessional motion about the $z$-axis.
Note that $\omega^\text{LLG}_\text{res}$ is renormalized by $\mathcal{G}^{s}_{\parallel}$ and 
$\mathcal{G}^{a}_{\parallel}$, accounting for the damping of the precession and the 
renormalization of $\gamma$, respectively (see Eq.~\eqref{gamma_effective}).

\subsection{Magnetocrystalline anisotropy}
\label{MAE_methods}
In absence of external magnetic fields, the gap opening in the spin excitation spectrum 
is uniquely due to the MAE (\textit{i.e.}\ anisotropy field) breaking the SU(2) rotational
symmetry~\cite{Manuel:2015}. The expression of $\omega^\text{LLG}_\text{res}$ in the LLG model 
provided in Eq.~\eqref{res_freq_llg} shows that the resonance frequency is proportional 
to $\mathcal{K}$, which can also be computed from ground state DFT calculations. 
Here, we discuss two different ground state methods to compute this quantity relying on 
the \textit{magnetic force theorem}~\cite{Oswald:1985,LKAG:1987,Daalderop:1990,Wang2:1996} 
and establish a connection with the MAE obtained using linear response theory, 
$\mathcal{K}_\text{susc}$. 

For uniaxial systems, the energy depends on the direction of the magnetic moment in a simple way:
$\mathcal{E}(\theta) \sim \mathcal{K}\cos^2\theta$, where $\theta$ is the angle that the magnetic moment makes with the $z$-axis, \textit{i.e.}
$\vec{M}/|\vec{M}| = \hat{n}(\theta,\varphi) = \left(\cos\varphi\sin\theta,\sin\varphi\sin\theta,\cos\theta\right)$.
To lowest order in the phenomenological expansion, the axial symmetry renders the energy independent of the azimuthal angle $\varphi$.
It follows that the magnitude of the MAE, $\mathcal{K}$, can be obtained from total energy differences for two different 
orientations of the magnetization (out-of-plane and in-plane). However, as $\mathcal{K}$ is at most a few meV's,
this approach requires very accurate total energies, which is computationally demanding.

Alternatively, one can use the \textit{magnetic force theorem}, which states that, if the changes in the
charge and magnetization densities accompanying the rotation of the spin moment are small, 
the total energy difference can be replaced by the band energy difference~\cite{Oswald:1985,LKAG:1987,Daalderop:1990}: 
\begin{equation}
  \mathcal{K}_\text{Band}= \mathcal{E}_\text{Band}(0^\circ) - \mathcal{E}_\text{Band}(90^\circ)\quad,
\label{K_eband_diff}
\end{equation}
where $\mathcal{E}_\text{Band}(\theta)$ is the band energy (sum of Kohn-Sham energy eigenvalues) of the system when 
the spin moment makes an angle $\theta$ with the $z$-axis:
\begin{align}
  \mathcal{E}_\text{Band}(\theta) = \int^{\varepsilon_\text{F}}_{-\infty}\!\!\!\!\!
            \dd\varepsilon\,(\varepsilon - \varepsilon_\text{F})\,\rho(\varepsilon;\theta) \quad .
            \label{e_band_form}
\end{align}
It contains the effect of the orientation of the magnetic moment through how the density of states
$\rho(\varepsilon;\theta)$ is modified upon its rotation.
This quantity is evaluated with a single non-self-consistent calculation, by orienting the exchange-correlation
magnetic field in the desired direction, $\vec{B}_{\text{xc}}(\vec{r}\,) = B_{\text{xc}}(\vec{r}\,)\,\hat{n}(\theta,\varphi)$
(rigid spin approximation~\cite{Samir_nc:2005}).

The MAE can also be evaluated from the magnetic torque, which corresponds to the first 
derivative of $\mathcal{E}_\text{Band}(\theta)$ with respect to the magnetic moment direction. Using 
the Hellman-Feynman theorem, the torque reads~\cite{Wang:1996,Julie:2006,Mankovsky:2009}:
\begin{equation}
\begin{split}
\mathcal{T}_\theta & = \frac{\partial \mathcal{E}_\text{Band}}{\partial\theta}\quad,\\ 
            & = \int\!\dd\vec{r}\;B_\text{xc}(\vec{r}\,)\,\frac{\partial\hat{n}(\theta,\varphi)}{\partial\theta}\cdot\vec{M}(\vec{r}\,;\theta) \quad.
\end{split}
\label{torque_formula_7}
\end{equation}
As for the band energy calculations, the torque is also obtained from a single non-self-consistent
calculation, under the same approximations.
It is non-vanishing if the output spin magnetization density $\vec{M}(\vec{r}\,;\theta)$ is not collinear
with the input magnetic moment direction.
Considering the expected form of the MAE for uniaxial symmetry, we should find
\begin{equation}
  \mathcal{T}_\theta = -\mathcal{K}_\text{Torque}\sin(2\theta) \quad .
\end{equation}
In practice, the torque can be evaluated at different angles $\theta$. In this work, two deviation angles 
have been considered: a large deviation angle with $\theta = 45^\circ$, as done in Ref.~\onlinecite{Wang:1996},
and a small one near self-consistency, $\theta = 5^\circ$. For such small deviations, 
one can connect $\mathcal{K}_\text{Torque}$ to the value of the 
MAE obtained from the magnetic susceptibility, $\mathcal{K}_\text{Susc}$. It is shown 
in Appendix~\ref{appendix_B} that when considering a small rotation angle $\theta$
and a constant magnitude of the exchange-correlation spin-splitting (frozen potential 
approximation),
\begin{equation}
\begin{split}
\mathcal{K}_\text{Susc} & = \frac{\mathcal{K}_{\text{Torque}}}{1 - 
\frac{4\chi^\text{KS}_{+-}(0)\mathcal{K}_\text{Susc}}{M^{2}_{z}}}\quad,\\
& \sim \frac{\mathcal{K}_{\text{Torque}}}{1 + \frac{B^\text{a}}{B_\text{xc}}}\quad.
\end{split}
\label{renormalized_MAE}
\end{equation} 
The previous expression shows that $\mathcal{K}_{\text{susc}}$ corresponds to the 
$\mathcal{K}_{\text{Torque}}$ (evaluated for a small deviation angle) renormalized by 
a prefactor $(1+\frac{B^\text{a}}{B_\text{xc}})^{-1}$. In fact, this result is 
similar to the renormalization observed for magnetic 
interactions computed from the magnetic susceptibility~\cite{Bruno:2002,Filipe:2017}. 
For the systems of interest ($3d$ and $4d$ transition metals impurities), $B^\text{a}$ 
is in the meV range while $B_\text{xc}$ is in the order of eV. Therefore, one expects 
small corrections due to this renormalization, and the two quantities should be in good agreement.

\section{Electronic structure of $3d$ and $4d$ impurities in B\lowercase{i}$_2$T\lowercase{e}$_3$ and B\lowercase{i}$_2$S\lowercase{e}$_3$}
\label{elec_str_3d_4d}
\begin{figure*}
\centering
\includegraphics[width=1.0\textwidth]{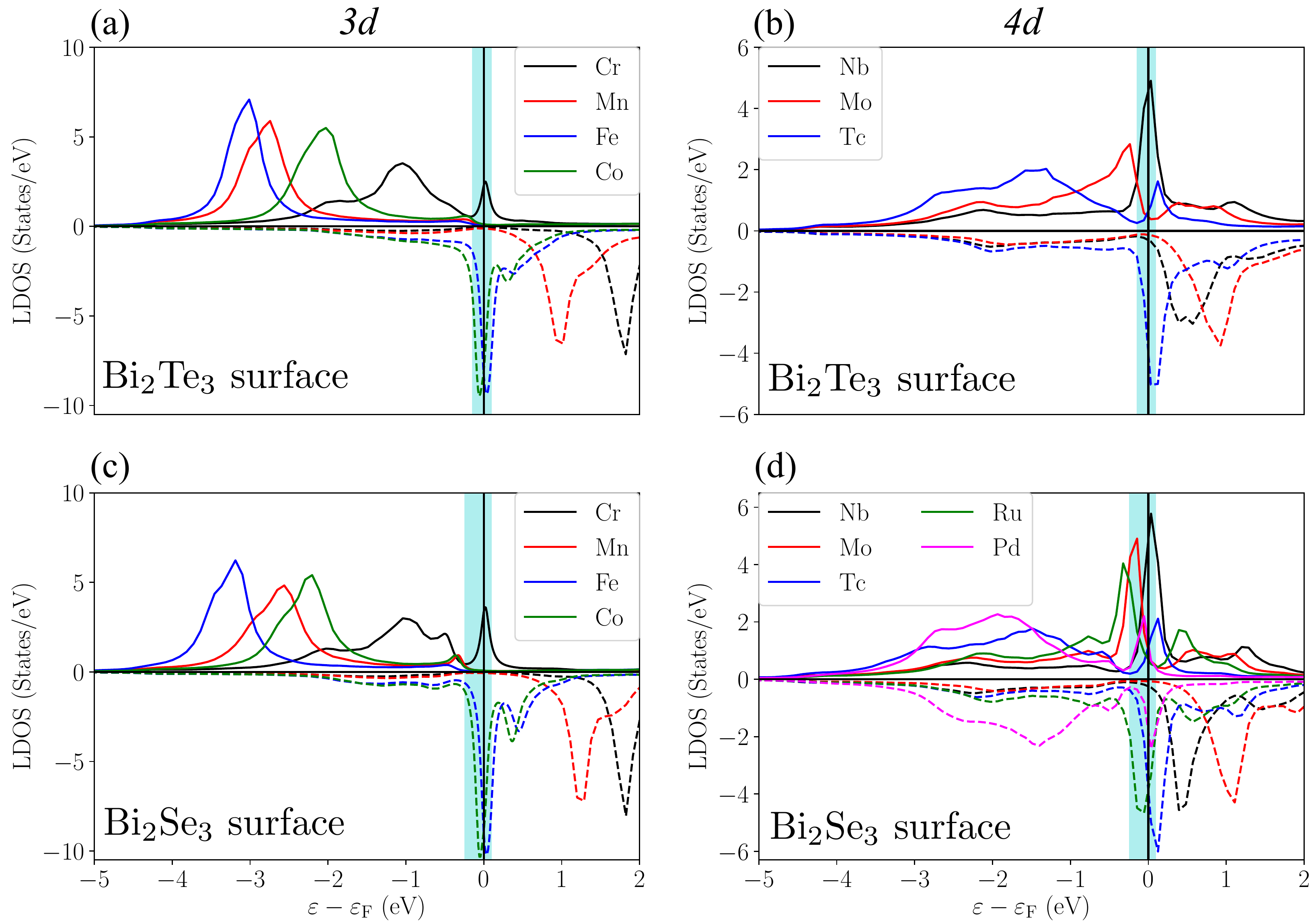}
\caption{\label{LDOS_3d_4d_Bi2Te3_Bi2Se3}
 Spin-resolved LDOS for $3d$ impurities (Cr, Mn, Fe and Co) and $4d$ impurities (Nb, Mo, Tc, Ru, Pd)
embedded in a Bi$_2$Te$_3$ (Bi$_2$Se$_3$) surface. (a) $3d$ in Bi$_2$Te$_3$, (b) $4d$ in 
Bi$_2$Te$_3$, (c) $3d$ in Bi$_2$Se$_3$ and (d) $4d$ in Bi$_2$Se$_3$. The full lines represent 
the majority-spin states, with dashed lines for the minority-spin ones. The energies 
are given with respect to the Fermi energy $\varepsilon_\text{F}$ and the energy window associated
with the bulk band gap is highlighted with light blue color.}
\end{figure*}

In this section, we briefly recap the discussion of the electronic structure and ground state 
properties of $3d$ impurities embedded in the Bi$_2$Te$_3$ (Bi$_2$Se$_3$) surface already addressed 
in Ref.~\onlinecite{Juba:2018}. Furthermore, we also consider $4d$ impurities which have a stronger 
hybridization with the host electrons compared to the $3d$ ones.
This information will be employed for the analysis of their dynamical properties, such as the Gilbert 
damping. The LDOS of $3d$ and $4d$ magnetic impurities embedded into Bi$_2$Te$_3$ and Bi$_2$Se$_3$(111) 
surfaces are shown in Fig.~\ref{LDOS_3d_4d_Bi2Te3_Bi2Se3}. The bulk band gap 
$(\Delta_\text{gap})$ is depicted in light blue --- with $\Delta_\text{gap}\approx 0.25$ eV 
for Bi$_2$Te$_3$ and $\Delta_\text{gap}\approx 0.35$ eV for Bi$_2$Se$_3$~\cite{Juba:2018}. 
We consider that the impurity spin moment is oriented perpendicularly to the surface ({\it i.e.} along the [111] direction). The full 
lines represent the majority spin channel $(\uparrow)$, while the dashed lines account 
for the the minority spin channel $(\downarrow)$. All the $3d$ and $4d$ 
impurities donate electrons to the host atoms (see Table~\ref{3d_4d_Bi2Te3_Bi2Se3_gs}). 
It can also be seen in Fig.~\ref{LDOS_3d_4d_Bi2Te3_Bi2Se3} that the spin splitting of the 
$4d$ impurities is weaker compared to the $3d$ ones, resulting in smaller spin moments, 
as listed in Table~\ref{3d_4d_Bi2Te3_Bi2Se3_gs}. This is attributed to the  Stoner parameter
being larger for $3d$ than for $4d$ elements~\cite{Janak:1977}.

\begin{table*}[b]
  \begin{center}
  \begin{ruledtabular}
  \begin{tabular}{c c c c c c c c c c c c}
     \multicolumn{2}{c}{}  & Cr    & Mn    & Fe    & Co    & Nb    & Mo    & Tc    & Ru      & Pd      \\
       \hline
        \multirow{2}{*}{$\mathcal{Q}$} & Bi$_2$Te$_3$ & 5.154 & 6.160 & 7.282 & 8.448 & 3.488 & 4.717 & 5.892 & 7.147   & 9.421   \\
        & Bi$_2$Se$_3$ & 4.841 & 5.863 & 6.963 & 8.136 & 3.077 & 4.316 & 5.474 & 6.734   & 9.041   \\
       \hline 
       \multirow{2}{*}{$M_\text{s}$} & Bi$_2$Te$_3$ & 3.843 & 4.412 & 3.395 & 2.108 & 1.097 & 2.678 & 2.493 & 0.000   & 0.000   \\
        & Bi$_2$Se$_3$  & 3.671 & 4.421 & 3.482 & 2.231 & 0.906 & 2.574 & 2.534 & 0.564   & 0.578   \\
       \hline
        \multirow{2}{*}{$M_\text{l}$} & Bi$_2$Te$_3$  & 0.065 & 0.050 & 0.260 & 0.883 &-0.143 &-0.004 & 0.202 & 0.000   & 0.000   \\
        & Bi$_2$Se$_3$  & 0.008 & 0.024 & 0.144 & 0.942 &-0.048 &-0.093 & 0.079 & 0.378   & 0.135   \\
  \end{tabular}
  \end{ruledtabular}
  \caption{Ground state properties of $3d$ and $4d$ impurities embedded in the Bi$_2$Te$_3$ 
  and Bi$_2$Se$_3$ surfaces including: the valence charge on the impurity $\mathcal{Q}$, 
  spin moment $M_\text{s}$ and orbital moment $M_\text{l}$. The spin and orbital moments are 
  given in units of $\mu_\text{B}$.
  \label{3d_4d_Bi2Te3_Bi2Se3_gs}}
  \end{center}
\end{table*}

All $3d$ elements except Cr display a completely filled majority-spin $d$-resonance. Mn and Cr 
have a nearly-empty minority-spin $d$-resonance, resulting in a large spin moment and a small orbital moment 
($M_\text{l}$). Fe and Co have a partially-filled minority-spin $d$-resonance, leading to higher values
for $M_\text{l}$, as shown in Table~\ref{3d_4d_Bi2Te3_Bi2Se3_gs}. The LDOS also reveals 
impurity-induced in-gap states near the Fermi energy, which arise from 
the hybridization with the bulk $sp$ states of Bi$_2$Te$_3$ (Bi$_2$Se$_3$)~\cite{Juba:2018}. 
When replacing the Bi$_2$Te$_3$ host by Bi$_2$Se$_3$, the valence charge and the 
spin moment are mildly affected, in contrast to the orbital moments which are 
considerably altered~\cite{Juba:2018}. 

For $4d$ impurities, both minority- and majority-spin $d$-resonances are partially occupied 
due to a weak spin-splitting. The LDOS is broader and flatter in comparison with the 
$3d$ ones, indicating a stronger hybridization with the host material, as 
the $4d$-orbitals are spatially more extended than the $3d$ ones, and so overlap more with the orbitals of the host. 
In the Bi$_2$Te$_3$ host, Nb, Mo and Tc are found to be magnetic, while Ru, Rh and Pd impurities
were found to be nonmagnetic. The analysis of the paramagnetic LDOS (not shown here)
reveals that, when moving in the periodic table from Tc towards Pd ({\it i.e.} adding electrons), 
the $4d$ peak is shifted to lower energies. This leads to a drastic decrease of the 
LDOS at $\varepsilon_\text{F}$ and makes the Stoner criterion unfulfilled. Nb has a less than 
half-filled $d$-shell, inducing an orbital moment anti-parallel to its spin moment, 
as shown in Table~\ref{3d_4d_Bi2Te3_Bi2Se3_gs}. For Mo and Tc, a half filled 
$d$-shell results in the highest values for $M_\text{s}$ between the $4d$ elements. These observations 
are in qualitative agreement with Hund's rules~\cite{Julen:2016}. In-gap states 
are also observed near $\varepsilon_\text{F}$, as for the $3d$ impurities.
Interestingly, in the Bi$_2$Se$_3$ host, Ru and Pd acquire a magnetic moment, while 
Rh remains nonmagnetic. Higher values of the LDOS at $\varepsilon_\text{F}$ 
compared to the Bi$_2$Te$_3$ host now satisfy the Stoner criterion for these 
elements. Pd is a rather peculiar case, since the increase of the LDOS at 
$\varepsilon_\text{F}$ is related to the presence of an in-gap state in the 
minority-spin LDOS, as shown in Fig.~\ref{LDOS_3d_4d_Bi2Te3_Bi2Se3}d.

The electronic structure, especially in the vicinity of the Fermi energy, 
governs the behaviour of the MAE and spin excitations of the system. 
In particular, the presence of $d$-resonances near $\varepsilon_\text{F}$ may result in inaccuracies 
in the computation of the MAE. Together with in-gap states, it can also induce 
high values of the Gilbert damping, as discussed in the next sections.

\section{Magnetocrystalline anisotropy of $3d$ and $4d$ impurities in 
B\lowercase{i}$_2$T\lowercase{e}$_3$ and B\lowercase{i}$_2$S\lowercase{e}$_3$}
\label{Mca_3d_4d_Bi2Te3}
\begin{figure*}
\centering
\includegraphics[width=1.0\textwidth]{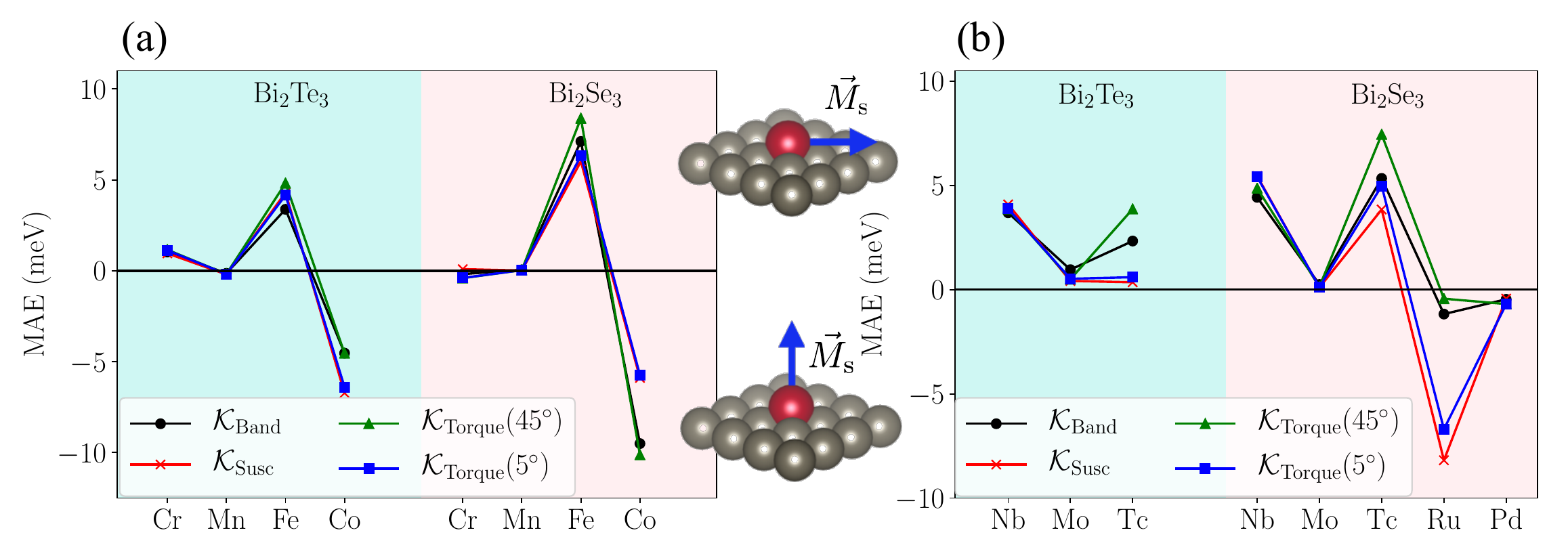}
 \caption{Comparison of the MAE for (a) $3d$ impurities and (b) $4d$ impurities, embedded
 in a Bi$_2$Te$_3$ and a Bi$_2$Se$_3$ surface. The black curve is obtained using the band energy differences 
 $(\mathcal{K}_\text{Band}$ [Eq.~\eqref{K_eband_diff}])
 (with a $90^{\circ}$ rotation of the spin moment). The red curve shows the MAE
 computed from the static part of the magnetic susceptibility $(\mathcal{K}_\text{Susc}$ 
 [Eq.~\eqref{fiting_inv_susc}]). The green and blue curves
 are obtained using the torque method at $45^{\circ}$ and $5^{\circ}$ $(\mathcal{K}_\text{Torque}(\theta) $[Eq.~\eqref{torque_formula_7}]), respectively. 
 Most of the impurities display an in-plane magnetic anisotropy ($\mathcal{K} > 0$).}
\label{MAE_3d_4d_methods}
\end{figure*}

We now investigate the MAE employing the different methods discussed in Sec.~\ref{MAE_methods}.
In our convention, a positive (negative) MAE stands for an in-plane (out-of-plane) easy-axis.  
In Fig.~\ref{MAE_3d_4d_methods}a, we show the evolution of the MAE for $3d$ impurities
embedded in Bi$_2$Te$_3$ and Bi$_2$Se$_3$, respectively. For every impurity, all the methods predict 
the same easy-axis. In the Bi$_2$Te$_3$ host, Cr and Fe present an in-plane magnetic anisotropy, 
while Mn and Co favor an out-of-plane orientation. The trend is mostly accounted for by
Bruno's formula~\cite{Bruno:1989}, where the MAE is given 
by the anisotropy of the orbital moment $(M_\text{l})$: $\mathcal{K} \propto \zeta^{2}\,(M_\text{l}^{x} - 
M_\text{l}^{z})$, with $\zeta$ being the spin-orbit interaction strength. 
Mn displays a small MAE, as it has a small orbital moment, while the large anisotropy energies
obtained for Fe and Co stem both from their large orbital moments and their substantial dependence on the spin orientation. 
However, the results obtained for the MAE of Cr do not agree with the predictions 
of Bruno's formula, since the MAE reaches$~\sim \SI{1}{\milli\electronvolt}$, despite 
a rather small anisotropy in the orbital moment of the adatom (see Table.~\ref{gs_2_hosts}). 
For the Bi$_2$Se$_3$ host, the anisotropy follows very similar trends in comparison with the Bi$_2$Te$_3$ 
case. Nonetheless, the easy axis of Cr switches from in-plane to out-of-plane, while the 
MAE of Fe and Co present a noticeable increase, as shown in Fig.~\ref{MAE_3d_4d_methods}a. These changes 
in the MAE are attributed to the modification of the ground state properties, particularly the 
orbital moments (as listed in Table~\ref{gs_2_hosts}), according to Bruno's formula.

\begin{table*}[b]
\begin{center}
\begin{ruledtabular}
\begin{tabular}{c c c c c c c c c c c c}
&                            & Cr      & Mn     & Fe     & Co      & Nb      & Mo     & Tc       & Ru          &   Pd   \\
\hline
\multirow{2}{*}{$\Delta \mathcal{Q}^{zx}$} & Bi$_2$Te$_3$ & -0.016  & 0.001  & -0.224 & -0.484  &  0.018  & 0.002  &  -0.287  & 0.000  &  0.000 \\
& Bi$_2$Se$_3$ & -0.001  & 0.000  & -0.320 & -0.583  & -0.004  & 0.001  &  -0.319  &-0.347  &  0.000 \\
\hline 
\multirow{2}{*}{$\Delta M_\text{s}^{zx}$} & Bi$_2$Te$_3$  & -0.016  & -0.001 & 0.224  & 0.483   &  0.0147 & -0.000 &   0.288  &  0.000 &  0.000 \\
& Bi$_2$Se$_3$  & -0.001  & -0.000 & 0.320  & 0.582   &  -0.009 &  0.001 &   0.286  &  0.320 & -0.003 \\
\hline
\multirow{2}{*}{$\Delta M_\text{l}^{zx}$} & Bi$_2$Te$_3$  &  0.019  & 0.003  & -0.323 & 0.484   &  -0.081  & -0.002 & -0.188  &  0.000 &  0.000 \\
& Bi$_2$Se$_3$  &  0.003  & 0.002  & -0.493 & 0.487   &  -0.261  &  0.003 & -0.284  &  0.285 &  0.008  \\
\end{tabular}
\end{ruledtabular}
\caption{Change in the valence charge of the impurity $\Delta \mathcal{Q}^{zx}$, spin moment $\Delta M^{zx}_\text{s}$ and orbital moment $\Delta M^{zx}_\text{l}$ for 
$3d$ and $4d$ impurities embedded in a Bi$_2$Te$_3$ and a Bi$_2$Se$_3$ surface, using the frozen
potential approximation. For Fe and Co, $\Delta\mathcal{Q}^{zx}$ and $\Delta M^{zx}_\text{s}$ are
relatively large, invalidating the use of the magnetic force theorem to compute the MAE.
\label{gs_2_hosts}}
\end{center}
\end{table*}

In Fig.~\ref{MAE_3d_4d_methods}b, we show the MAE of $4d$ impurities embedded in Bi$_2$Te$_3$ 
and Bi$_2$Se$_3$ computed with the different approaches outlined in Section~\ref{MAE_methods}. 
For the Bi$_2$Te$_3$ case, all the impurities (Nb, Mo and Tc) display an in-plane easy-axis. Nb displays a large 
MAE, while Mo and Tc have a rather small one (with the exception of $\mathcal{K}_\text{Torque}(45^{\circ})$
and $\mathcal{K}_\text{Band}$).
For Mo, the small MAE correlates with its small orbital moment. In the Bi$_2$Se$_3$ host, 
Nb, Mo, and Tc are characterized by an in-plane easy-axis as well. Note that, due to a strong
hybridization with the host (broad LDOS in Fig.~\ref{LDOS_3d_4d_Bi2Te3_Bi2Se3}b 
and d), the MAE of Tc is drastically affected by the surrounding environment. Ru and Pd acquire 
a magnetic moment in Bi$_2$Se$_3$ displaying an out-of-plane easy-axis. Particularly, Ru
displays a very large MAE in comparison with the rest of the $4d$ elements. 

We now focus on the reasons why different methods may provide contrasting values for the MAE 
(see Fig.~\ref{MAE_3d_4d_methods}). The origin of these divergences can be traced back to 
the features of the electronic structure at the impurity site. Fig.~\ref{MAE_3d_4d_methods}a shows 
that the obtained MAE energies of Fe and Co can be separated in two groups, according to the method
used to compute them: One for large angle methods, including the band energy differences $(\mathcal{K}_\text{Band}$ [Eq.~\eqref{K_eband_diff}]) 
and the torque method at $45^{\circ}$ $(\mathcal{K}_\text{Torque}(45^{\circ})$ [Eq.~\eqref{torque_formula_7}]); and the other for
small perturbations, encompassing the torque method at $5^{\circ}$ $(\mathcal{K}_\text{Torque}(5^{\circ}) $[Eq.~\eqref{torque_formula_7}])
and linear response theory $(\mathcal{K}_\text{Susc}$ [Eq.~\eqref{fiting_inv_susc}]). The results from the two methods in each group
are in good agreement with each other, but the results from one group do not agree with those from the other.
This can be understood via Table~\ref{gs_2_hosts}, 
which lists the change in the ground state properties of the impurity upon $90^{\circ}$ rotation
of the spin moment ($z \rightarrow x$ axis), in a frozen potential calculation. 
There is a large variation in the valence charge and in the spin moment of Fe and Co in comparison 
to Cr and Mn, owing to the change in the position of the $3d$ peak in the minority spin channel in the vicinity of \
$\varepsilon_\text{F}$ (see Fig.~\ref{LDOS_3d_4d_Bi2Te3_Bi2Se3}a 
and~\ref{LDOS_3d_4d_Bi2Te3_Bi2Se3}c). This violates the assumptions 
justifying the magnetic force theorem (in the frozen potential approximation), as previously observed 
in Ref.~\onlinecite{Pick:2003} for Co adatoms deposited on a Cu(111) surface.
The disagreement between the different methods for Tc and Ru observed in Fig.~\ref{MAE_3d_4d_methods}b
is attributed to a high occupation at $\varepsilon_\text{F}$ as well (see Fig.~\ref{LDOS_3d_4d_Bi2Te3_Bi2Se3}b and~\ref{LDOS_3d_4d_Bi2Te3_Bi2Se3}d). An exception occurs for Nb, where good agreement 
between the different methods is observed. In this case, the high LDOS at $\varepsilon_\text{F}$
is due to the majority spin states, which are weakly affected by the spin rotation. 

The previous analysis indicates that, if a high density of electronic states is present at $\varepsilon_\text{F}$ 
(Fe, Co, Tc and Ru), a large rotation angle may lead to large changes in the charge density and invalidate the 
use of the \textit{magnetic force theorem} in combination with the frozen potential approximation.
Therefore, a small deviation angle, for which the system remains near self-consistency, should be
considered. This can be achieved through the torque method or the magnetic susceptibility. The MAE obtained in these cases $(\mathcal{K}_{\text{Torque}}(5^{\circ})$ and $\mathcal{K}_{\text{Susc}})$ should be comparable
with the one extracted for inelastic scanning tunneling spectroscopy measurements, since in 
such experiments the deviation of the magnetic moment from the easy-axis are rather small.


\section{Spin excitations of $3d$ and $4d$ impurities in B\lowercase{i}$_2$T\lowercase{e}$_3$ and B\lowercase{i}$_2$S\lowercase{e}$_3$}
\label{SE_3d_4d_allhosts}

\begin{figure*}
\includegraphics[width=1.0\textwidth]{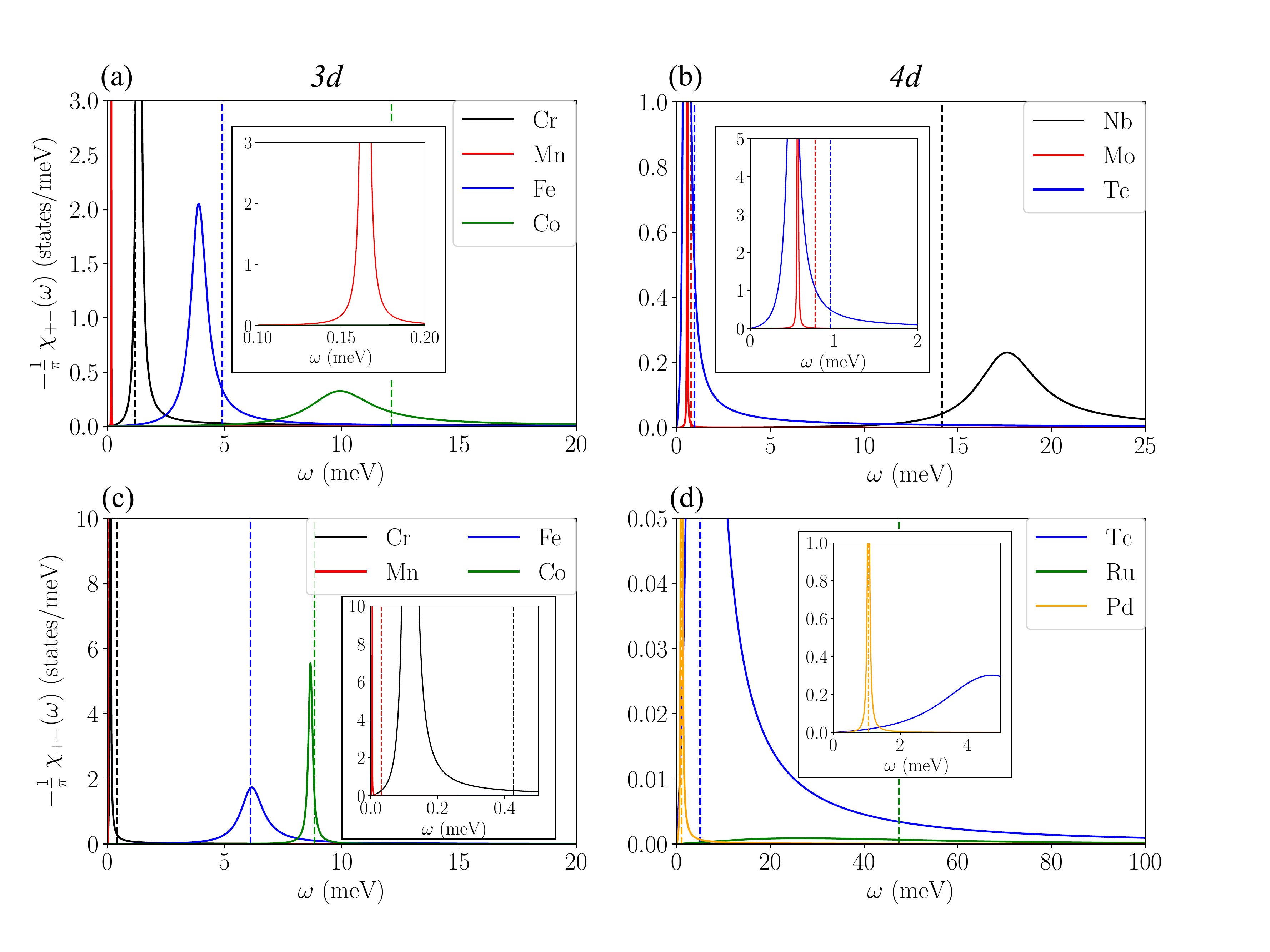}
\caption{Density of states of transverse spin excitations for magnetic impurities. 
The panels show the results for (a) $3d$ and (b) $4d$ impurities embedded in Bi$_2$Te$_3$, 
and (c) $3d$ and (d) $4d$ impurities embedded in Bi$_2$Se$_3$.
They present an almost-Lorentzian, with resonances located at the excitation energies of the system.
The dashed lines mark the resonance frequency without dynamical corrections, 
$\omega^{0}_\text{res} =- \frac{2\gamma\mathcal{K}_\text{Susc}}{M_\text{s}}$.
For Mn, Co, Ru and Pd, $\chi_{-+}(\omega)$ is plotted instead, to account for their easy-plane MAE.}
\label{SE_3d_4d_2hosts}
\end{figure*}

In Sec.~\ref{elec_str_3d_4d}, we addressed the ground state properties of $3d$ and $4d$ 
impurities embedded in Bi$_2$Te$_3$ and Bi$_2$Se$_3$. Here, we investigate their spin 
dynamics, relate it to the MAE obtained in Sec.~\ref{Mca_3d_4d_Bi2Te3}, and study the 
possibility of exciting and manipulating these impurities with time-dependent 
external magnetic fields. We focus on the transverse spin excitations encoded in the
dynamical magnetic susceptibility, which have been observed experimentally for magnetic 
impurities on nonmagnetic surfaces by means of ISTS measurements~\cite{Heinrich:2004,Hirjibehedin:2006,Hirjibehedin:2007,Balashov:2009,Alex:2011,Khajetoorians:2013}. 
In these experiments, the spin excitations yield a step in the differential tunneling conductance 
at well-defined energies.

We show in Fig.~\ref{SE_3d_4d_2hosts} the imaginary part of $\chi_{+-}(\omega)$ (\textit{i.e.}\ 
the density of states of the magnetic excitations) as function of the frequency of the external field for both $3d$ and 
$4d$ impurities embedded in Bi$_2$Te$_3$ and Bi$_2$Se$_3$. Only the response of the magnetic 
impurities is considered, since the induced moments in the surrounding (host) atoms are 
rather small. Nonetheless, their contribution is accounted for when computing the transverse 
exchange-correlation kernel $K_{\perp}^\text{xc}$ at the impurity site via the spin-splitting 
sum rule~\cite{Samir:2010,Manuel:2015}. The LLG parameters obtained by fitting the data 
to Eq.~\eqref{fiting_inv_susc} are given in Table~\ref{SE_parameters_Bi2Te3_Bi2Se3}.
\begin{table*}[b!]
\begin{center}
\begin{ruledtabular}
\begin{tabular}{c c c c c c c c c c c c}
            &                & Cr      & Mn     & Fe     & Co      & Nb      & Mo     & Tc       & Ru          &   Pd   \\
\hline
\multirow{2}{*}{$M_\text{s}$} & Bi$_2$Te$_3$       &  3.844  &  4.412  & 3.395  & 2.109  &  1.097   & 2.678   &  2.493     & ---     & ---  \\
 & Bi$_2$Se$_3$                                    &  3.671  &  4.421  & 3.482  & 2.231  &  0.906   & 2.574   & 2.534      &  0.564  & 0.578  \\
\hline 
\multirow{2}{*}{$\mathcal{G}^{s}_{\parallel}$} & Bi$_2$Te$_3$  &  0.019  &  0.000 & 0.143  & 0.164   &  0.053   &  0.000   &  0.172  & ---   & ---  \\
                                               & Bi$_2$Se$_3$  &  0.037  &  0.000 & 0.112  & 0.012   &  0.003   &  0.000   &  0.512  & 0.852 & 0.094  \\
\hline
\multirow{2}{*}{$\mathcal{G}^{a}_{\parallel}$} & Bi$_2$Te$_3$  & -0.245  & 0.109  & 0.286  &  0.274  &  -0.087   &  0.096  &  0.099  &  ---    & --- \\
                                               & Bi$_2$Se$_3$  & -0.153  & 0.101  & 0.125  &  0.196  &  -0.021   &  0.134  &  0.081  & -0.396  & 1.824  \\
\hline
\multirow{2}{*}{$\omega_\text{c}$}   & Bi$_2$Te$_3$ &  77.68   &   3439  &   135.7   &    277.4  &  21.91   &  224.5  &   31.64  &   ---    &  ---   \\
                                     & Bi$_2$Se$_3$ &  283.2   &   1340  &   100.4   &    73.37  &  2.784   &  403.5  &   4.481  &  10.11   & 437.0   \\
\hline
\multirow{2}{*}{$\eta_\text{c}$}     & Bi$_2$Te$_3$ &  7.154   &  298.3   &  65.66    &  38.39    &   30.36  &  752.2  &  234.4   & ---      &  ---       \\
                                     & Bi$_2$Se$_3$ &  30.97   &  17820   &  76.31    &  40.19    &   8.703  &  171.5  &  84.93   &  341.8   &  502.5     \\
                                   \hline 
\multirow{2}{*}{$\mathcal{K}_\text{Susc}$} & Bi$_2$Te$_3$      &  0.959   & -0.201    &  4.302    &   -6.725 &  4.091   &  0.417  & 0.353    &  ---     &  --- \\
                                           & Bi$_2$Se$_3$      &  0.090   &  0.005    &  6.019    &   -5.894 &  5.453   &  0.102  & 3.845    & -8.178   &  -0.431 \\
\hline
\multirow{2}{*}{$\omega^\text{LLG}_\text{res}$} & Bi$_2$Te$_3$ &  1.322 & 0.164 & 3.917 & 9.926 & 16.31 & 0.568 & 0.509 & --- & --- \\
                                                & Bi$_2$Se$_3$ &  0.115 & 0.004 & 6.113 & 8.833 & 24.08 & 0.158 & 5.073 & 55.49 & 1.055 \\
\hline
\multirow{2}{*}{$\dfrac{\omega^\text{LLG}_\text{res}}{\omega_\text{c}}$} & Bi$_2$Te$_3$ &  0.017 & 0.000 & 0.029 & 0.036 & 0.744 & 0.003 & 0.016 & ---   & --- \\
                                                                         & Bi$_2$Se$_3$ &  0.000 & 0.000 & 0.063 & 0.125 & 8.836 & 0.000 & 1.132 & 5.487 & 0.002\\
\end{tabular}
\end{ruledtabular}
\caption{LLG parameters for $3d$ and $4d$ impurities embedded in the surface of Bi$_2$Te$_3$ (Bi$_2$Se$_3$), 
obtained by fitting the TDDFT dynamical susceptibility data to Eq.~\eqref{fiting_inv_susc}.
$M_\text{s}$ is the spin moment of the impurity. $\mathcal{G}^{s}_{\parallel}$ is the symmetric 
part and $\mathcal{G}^{a}_{\parallel}$ is the antisymmetric part of the damping tensor, both unitless. 
$\mathcal{K}_\text{Susc}$ is the MAE obtained from the magnetic susceptibility, in meV. 
$\omega^\text{LLG}_\text{res}$ is the resonance frequency without including nutation, in meV, 
as defined in Eq.~\eqref{res_freq_llg}. A large ratio between $\omega^\text{LLG}_\text{res}$ 
and $\omega_\text{c} = \frac{\mathcal{G}^{a}_{\parallel}}{\mathcal{I}^\text{s}_{\parallel}}$ 
indicates that the nutation makes a substantial contribution to $\omega_\text{res}$, while 
$\eta_\text{c}=\frac{\mathcal{G}^{s}_{\parallel}}{\mathcal{I}^\text{a}_{\parallel}}$ provides 
information on the contribution of the nutation to the damping of the spin excitation.
Ru and Pd in Bi$_2$Te$_3$ were found to be nonmagnetic, so the corresponding entries are 
marked with a dash.
\label{SE_parameters_Bi2Te3_Bi2Se3}}
\end{center}
\end{table*}
As depicted in Fig.~\ref{SE_3d_4d_2hosts}, $\text{Im}\,\chi_{+-}(\omega)$ has a Lorentzian-like 
shape, and the resonance frequency $(\omega_\text{res})$ is finite even in absence 
of an external magnetic field. This resonance arises from the MAE, which breaks the SU(2) 
rotational symmetry (\textit{i.e.}~no Goldstone mode), as explained previously in Sec.~\ref{MAE_methods}. 
The highest resonance frequencies are obtained for Nb and Ru due to their strong anisotropy 
combined with a small magnetic moment complying with Eq.~\eqref{res_freq_llg}, while 
the smallest value of $\omega_\text{res}$ is obtained for Mn impurities in Bi$_2$Se$_3$. 
The dashed lines in Fig.~\ref{SE_3d_4d_2hosts} represent the resonance position obtained 
neglecting dynamical corrections in Eq.~\eqref{res_freq_llg}, leading to the estimate
$\omega^{0}_\text{res} = -\frac{2\gamma\mathcal{K}_\text{Susc}}{M_\text{s}}$ (with $\gamma = 2$ and 
$\underline{\mathcal{G}} = 0$)~\cite{Manuel:2015}.  There is a qualitative agreement between 
$\omega^{0}_\text{res}$ and the resonance position extracted from the spin excitation spectra, 
$\omega_\text{res}$, including damping and nutation. Nonetheless, their values are quantitatively 
different, illustrating that dynamical corrections can be of crucial importance for an 
accurate determination of the resonance frequency. 

\begin{figure}
\centering
\includegraphics[width=0.50\textwidth]{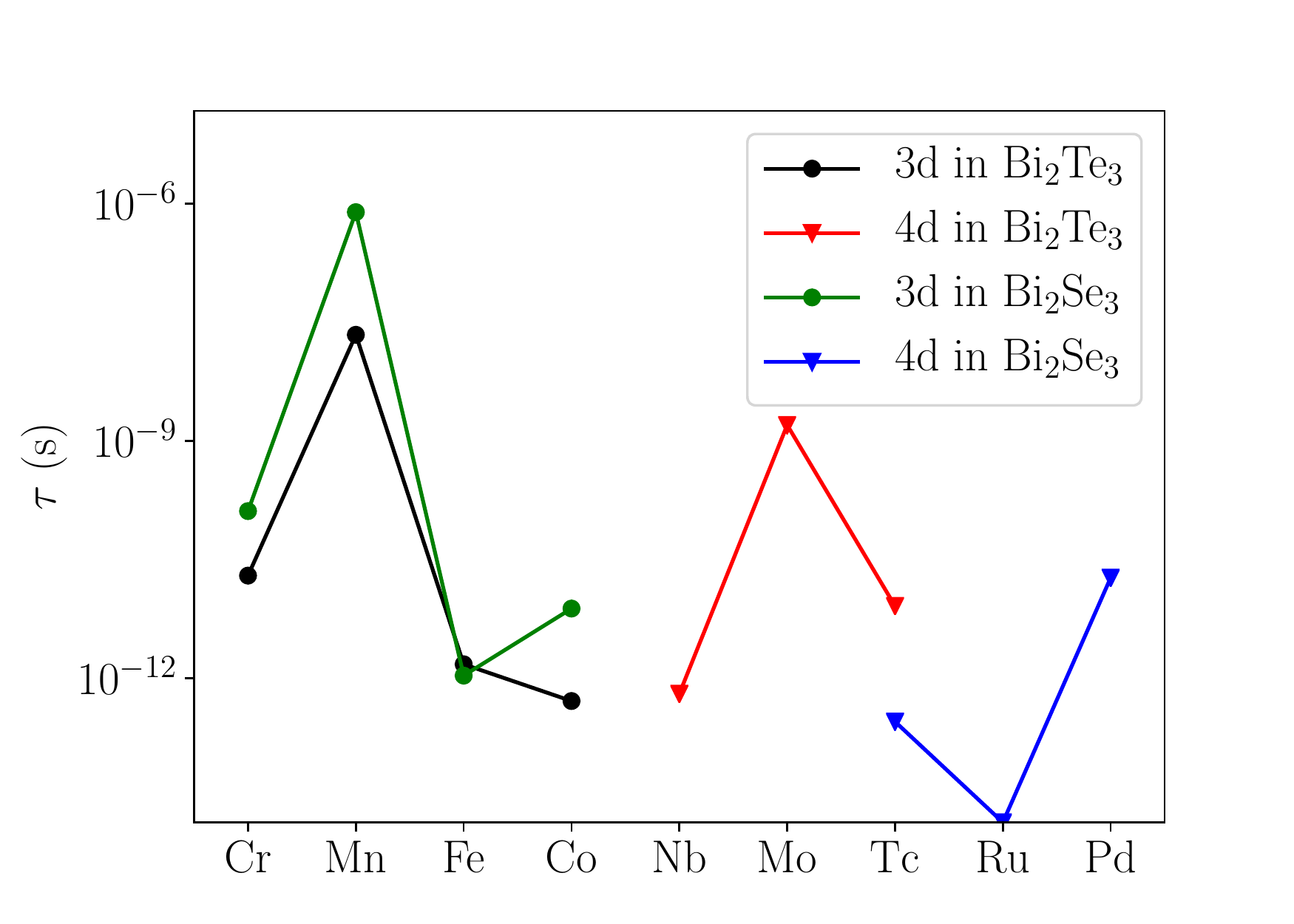}
\caption{Excitation lifetime of $3d$ and $4d$ magnetic impurities embedded in 
Bi$_{2}$Te$_3$ and Bi$_2$Se$_3$. Note that the lifetime axis is on a logarithmic 
scale. The highest excitation lifetime is obtained for Mn in Bi$_2$Se$_3$ and 
reaches microseconds, while the lowest one is obtained for Ru. Elements without 
data were found to be nonmagnetic in the respective hosts.}
\label{SE_liftimes}
\end{figure}

Another quantity which is strongly dependent on the nature of the impurity and the host
is the full width at half maximum (FWHM) $\Gamma$. This quantity is proportional to the 
symmetric part of the Gilbert damping tensor ($\mathcal{G}^{s}_{\parallel}$)  and 
provides information about the lifetime of the excitations~\cite{Julen:2017} as $\tau = \frac{2}{\Gamma}$. 
This lifetime ranges from \textit{picoseconds} (comparable to
lifetimes obtained at metallic surfaces~\cite{Manuel:2015,Julen:2017}) to 
very high values reaching \textit{microseconds} for Mn in Bi$_2$Se$_3$ as shown in Fig.~\ref{SE_liftimes}.
Furthermore, the values of $\mathcal{G}^{s}_{\parallel}$, 
shown in Table~\ref{SE_parameters_Bi2Te3_Bi2Se3}, can be interpreted in 
terms of the LDOS at $\varepsilon_\text{F}$, since $\mathcal{G}^\text{s}_{\parallel} 
\propto n^{\downarrow}(\varepsilon_\text{F})\,n^{\uparrow}(\varepsilon_\text{F})$ 
(where $n^{\downarrow}(\varepsilon)$ and $n^{\uparrow} (\varepsilon)$ represents the 
LDOS of the minority and majority spin channels, respectively)~\cite{Samir:2015}. 
The highest values of $\mathcal{G}^{s}_{\parallel}$ are obtained for Ru, which 
coincide the lowest excitation lifetime as displayed in Fig.~\ref{SE_liftimes}.  
The anti-symmetric part of the Gilbert damping tensor $\mathcal{G}^{a}_{\parallel}$ is also 
displayed in Table~\ref{SE_parameters_Bi2Te3_Bi2Se3}. It accounts for the 
renormalization of the gyromagnetic ratio, $\gamma_\text{eff} = \frac{\gamma}{1
+\mathcal{G}^{a}_{\parallel}}$ (see Appendix~\ref{Append_A}). This renormalization is 
attributed to the presence of a finite LDOS at $\varepsilon_\text{F}$ as well~\cite{Samir:2015}. 
$\mathcal{G}^{a}_{\parallel}$ is negative for Cr, Nb and Ru indicating an enhancement 
of the gyromagnetic ratio ({\it i.e.} $\gamma_\text{eff} > 2$), while $\gamma_\text{eff} < 2$ 
for the remaining impurities. Note that the spin excitation spectra of Nb and Mo impurities 
in Bi$_2$Se$_3$ is not shown in Fig.~\ref{SE_3d_4d_2hosts}, since for these elements 
the Taylor expansion shown in Eq.~\eqref{Taylor_exp_KSsusc} fails due to contributions 
from higher order terms in frequency becoming too large.

The importance of the nutation can be estimated from the real part of the denominator 
of Eq.~\eqref{chi_pm_llg}. Both damping and nutation terms, $\mathcal{G}^{a}_{\parallel}\omega$ 
and $\mathcal{I}^{s}_{\parallel}\omega^{2}$, contribute to the resonance. 
When it occurs at frequencies higher than $\omega_\text{c} = \frac{\mathcal{G}^{a}_{\parallel}}
{\mathcal{I}^\text{s}_{\parallel}}$, $\omega_\text{res}$ can be substantially affected 
by the nutation. The ratio between $\omega^\text{LLG}_\text{res}$ obtained using 
Eq.~\eqref{res_freq_llg} (without including nutation) and $\omega_\text{c}$ (shown 
in Table~\ref{SE_parameters_Bi2Te3_Bi2Se3}) is employed to evaluate the importance of this 
contribution. The symmetric parts of the Gilbert damping and nutation tensors can be also 
related via~\cite{Ciornei:2011,Bottcher:2012} $\mathcal{I}^\text{s}_{\parallel} \propto 
\mathcal{G}^\text{s}_{\parallel}$, \textit{i.e.} the damping and nutation coefficients 
are proportional. The ratio $\omega_c$ is fairly small for the majority of the elements, 
indicating that nutation has no significant impact on the resonant spin precession. 
However, for some elements such as Nb and Tc (in Bi$_2$Se$_3$) the nutation leads to 
a shift of $\sim 1.3$ and $0.4$ meV in the resonance frequency, respectively. Finally, 
the most striking element is once again Ru, with a shift of the resonance frequency from 
$\omega^\text{LLG}_\text{res}=55.49$ to $\omega_\text{res}=25.52$ due to the nutation.

\section{Surface and bulk spin dynamics}
\label{spin_dyn_bulk_VS_surface}

We now compare different cases of $3d$ and $4d$ magnetic impurities embedded in a surface 
and in a bulk inversion symmetric Bi$_2$Te$_3$ (\textit{i.e.} insulating phase with 
no topological surface state). This enables us to disentangle the surface and bulk 
contributions to the spin dynamics. The analysis of the ground state properties of 
the $3d$ impurities embedded in bulk Bi$_2$Te$_3$ is given in Ref.~\onlinecite{Juba:2018}. 
The impurity-induced electronic in-gap states are also present in $4d$ impurities embedded in bulk Bi$_2$Te$_3$.
The LLG parameters obtained in the bulk (denoted with a subscript ``b'') and at the surface 
(denoted with a subscript ``s'') are displayed in Table~\ref{3d_SE_bulk_surface}. 
\begin{table*}[tb!]
\begin{center}
\begin{ruledtabular}
\begin{tabular}{c c c c c c c c c}
 & $M_\text{s}$ & $\mathcal{G}^{s}_{\parallel}$ & $\mathcal{G}^{a}_{\parallel}$ 
&  $\omega_\text{c}$  &  $\eta_\text{c}$  & $\mathcal{K}_\text{Susc}$ 
&  $\omega^\text{LLG}_\text{res}$ &  $\dfrac{\omega^\text{LLG}_\text{res}}{\omega_\text{c}}$     \\
\hline
Cr$_\text{s}$  & 3.844   &  0.018  & -0.245  & 77.68   &  7.154  &  0.959   &  1.322  &  0.017  \\
Cr$_\text{b}$  & 3.823   &  0.004  & -0.215  & 332.6   &  47.48  & -0.824   &  1.090  &  0.003  \\
\hline 
Mn$_\text{s}$  & 4.412   &  0.000  &  0.109  &  3439  & 298.4 &  -0.201  &  0.164  &  0.000  \\
Mn$_\text{b}$  & 4.335   &  0.000  &  0.118  &  860.7  & 590.4   &  -0.216  &  0.178  &  0.000  \\
\hline
Fe$_\text{s}$  & 3.395   &  0.143  &  0.286  &  135.7  &  65.66  &   4.302  &  3.917  &  0.029  \\
Fe$_\text{b}$  & 3.294   &  0.045  &  0.234  & 58.98   & 20.87   &   3.055  &  3.004  &  0.053  \\
\hline
Co$_\text{s}$  & 2.109   &  0.164  &  0.274  & 277.4   & 38.39   &  -6.725  &  9.926  &  0.037  \\
Co$_\text{b}$  & 1.977   &  0.307  & -0.011  &  1.015  & 56.09   &  -2.168  &  4.237  &  4.174  \\
\hline
Nb$_\text{s}$  & 1.097   &  0.053  & -0.087  &  21.91  &   30.36 &  4.091   & 16.31  &  0.769  \\
Nb$_\text{b}$  & 0.740   &  0.314  &  0.049  &  10.59  & 488.5   &  1.028   &  5.074  &  0.479  \\
\hline
Mo$_\text{s}$  & 2.678   &  0.000  &  0.096  &  224.5  &  752.2  &  0.417   &  0.568  &  0.003  \\
Mo$_\text{b}$  & 2.527   &  0.012  &  0.151  & 323.9   & 1083   &  0.454   &  0.624  &  0.002  \\
\hline
Tc$_\text{s}$  & 2.493   &  0.172  &  0.099  &  31.64  & 234.4 &  0.353   &  0.509  &  0.016  \\
Tc$_\text{b}$  & 2.057   &  0.059  &  0.072  & 12.67   &  29.32  &  0.755   &  1.368  &  0.111 \\
\end{tabular}
\end{ruledtabular}
\caption{LLG parameters for $3d$ and $4d$ impurities embedded in the surface (subscript s) and 
in the bulk (subscript b) of Bi$_2$Te$_3$, obtained by fitting the TDDFT dynamical susceptibility 
data to Eq.~\eqref{fiting_inv_susc}. $M_\text{s}$ is the spin moment of the impurity. 
$\mathcal{G}^{s}_{\parallel}$ is the symmetric part and $\mathcal{G}^{a}_{\parallel}$ is the 
antisymmetric part of the damping tensor, both unitless. $\mathcal{K}_\text{Susc}$ is the MAE obtained 
from the magnetic susceptibility, in meV. $\omega^\text{LLG}_\text{res}$ is the resonance 
frequency without including nutation, in meV, as defined in Eq.~\eqref{res_freq_llg}. A large ratio between $\omega^\text{LLG}_\text{res}$ and $\omega_\text{c} = \frac{\mathcal{G}^{a}_{\parallel}}{\mathcal{I}^\text{s}_{\parallel}}$ indicates that the nutation makes a substantial contribution to $\omega_\text{res}$, while $\eta_\text{c}=\frac{\mathcal{G}^{s}_{\parallel}}
{\mathcal{I}^\text{a}_{\parallel}}$ provides information on the contribution of the nutation to the damping of the spin excitation.
The MAE and the Gilbert damping are considerably affected when going from 
surface to bulk. The largest changes occur in the case of the Co impurity.}
\label{3d_SE_bulk_surface}
\end{center}
\end{table*}
With the exception of Mn, the MAE obtained from the susceptibility differs 
considerably between the bulk and surface cases --- Cr even has its easy-axis 
switched. The overall change in the MAE is a decrease from the surface to the 
bulk cases. The immediate environment of the embedded impurities is the same 
in bulk and at surface. However, for the bulk case, the missing contribution 
of the surface state leads to modifications in the electronic structure, altering 
the virtual bound and the in-gap states~\cite{Juba:2018}. This results in a reduction of the 
MAE.
The spectral weight at the Fermi level is also affected leading to a modification of the damping parameter~\cite{Samir:2015}. For Cr, Fe and Tc, $\mathcal{G}^{s}_{\parallel}$ decreases, while for Co, Nb and Mo, it increases.
$\mathcal{G}^{a}_{\parallel}$ follows similar trends as in the surface case. 
Co and Nb are the exception since $\mathcal{G}^{a}_{\parallel}$ switches sign, 
resulting in a change of $\gamma_\text{eff}$. 
The nutation is negligible for most of elements, except for Nb and Co --- for the latter, it 
leads to a noticeable shift of the resonance frequency from $\omega^\text{LLG}_\text{res} 
= 4.24$ meV to $\omega_\text{res} = 4.68$ meV. In summary, Co and Nb
impurities are very sensitive to the the presence of the surface state,
where the impurity states display rather different behaviours in the bulk
and at the surface leading to a different spin excitational nature.
In contrast, Mn impurities have a similar behavior in the bulk and at the surface, showing that 
the topological surface state plays a negligible role for their spin dynamics.

\section{Conclusions}
In this paper, we employed a first-principles approach for the investigation of the 
spin excitation spectra of $3d$ and $4d$ impurities embedded in two prototypical topological 
insulators, namely Bi$_2$Te$_3$ and Bi$_2$Se$_3$. The simulations were carried out using 
linear response TD-DFT in the framework of the KKR-GF method, suitable for computing the properties
of spin excitations at the nanoscale. A mapping onto a generalized LLG model allowed 
to extract from first-principles the MAE and transversal components of the Gilbert damping 
and nutation tensor. The obtained values of the MAE were then compared systematically to
the ones obtained using the torque method and band energy differences, that rely on the
magnetic force theorem and the frozen potential approximation.  

All the considered $3d$ impurities acquire a finite magnetic moment in both hosts, while 
the strong hybridization of the $4d$ impurities with the host states makes them more 
sensitive to the surrounding environment. For instance, Ru and Pd were found to be nonmagnetic 
in Bi$_2$Te$_3$ but became magnetic in Bi$_2$Se$_3$. Furthermore, and independently from 
nature of the orbitals ($3d$ or $4d$), large rotation angles result in significant changes in the electronic 
properties when a high electronic density of states is found at the Fermi energy, invalidating the 
assumptions made to invoke the magnetic force theorem. The MAE must be then
computed employing perturbative methods such as linear response theory or the torque method
with small deviation angles. The MAE obtained using linear response theory is found to 
coincide with the one computed from the torque method differing only by a negligible
renormalization factor. 

The spin excitation spectra of the impurities displays diverse trends. When the impurity virtual bound 
states or in-gap states are located away from the Fermi energy, the Gilbert damping 
is rather low and the lifetime of the excitation reaches high values compared to the 
ones observed in metallic hosts~\cite{Manuel:2015,Julen:2017}. The most striking
example is a Mn impurity in Bi$_2$Se$_3$, where the lifetime reaches \textit{microseconds}. 
A contrasting situation is observed for Ru, which displays a flat excitation resonance 
in conjunction with a low lifetime. Moreover, we found that nutation effects can be 
important and lead to important shifts of the resonance frequency for some elements 
such as Nb, Tc and Ru. Moreover, we examined the contribution of the surface state 
to the spin dynamics by comparing the LLG parameters of the impurities embedded in the surface 
with those of impurities embedded in the bulk.
For Co and Nb impurities, it was found that the topological surface state has a drastic impact on 
the dynamics via the spectral shift of the impurity-induced electronic in-gap states, while 
it plays a minor role for Mn impurities. 

We provided a systematic investigation of the spin dynamics of $3d$ and $4d$ 
impurities embedded in topologically insulating hosts. The results obtained for 
excitation lifetimes of some specific impurities (Mn) provide insights on the dual (metal and insulator)
nature of these materials. In addition to that, the MAE computed employing perturbative 
methods such as the linear response can be compared to the one extracted from ISTS measurements. 
Finally, several aspects remain to be uncovered from first principles: the zero-point spin fluctuations~\cite{Julen:2016}
of these impurities, which can be accessed via the dynamical magnetic susceptibility, 
as well the spin dynamics of magnetic nanoclusters or full magnetic layers
deposited on topological insulators.

\textbf{Acknowledgements} We thank Dr. Julen Iba\~nez-Azpiroz for fruitful discussions. This work 
was supported by the European Research Council (ERC) under the European Union's Horizon 
2020 research and innovation programme (ERC-consolidator grant 681405 DYNASORE).
We gratefully acknowledge the computing time granted by the JARA-HPC Vergabegremium 
and VSR commission on the supercomputer JURECA at Forschungszentrum J\"ulich.

\appendix
\section{Phenomenological parameters from the generalized Landau-Lifshitz-Gilbert equation}
\label{Append_A}
In this Appendix, we provide the explicit forms of the phenomenological quantities 
(anisotropy field, damping and nutation tensors) discussed in section~\ref{LLG_section}. 
First, we establish a connection between the anisotropy field $\vec{B}^\text{a}$ and 
the magnetocrystalline anisotropy using the phenomenological form of the band energy 
$\mathcal{E}_\text{Band}$. For ease of connection with the LLG, we present the derivation
using a vector formalism. For systems with uniaxial symmetry, the expansion of the 
band energy in terms of the magnetization up to second order reads~\cite{Wang:1996}
\begin{equation}
\mathcal{E}_\text{Band} = \mathcal{E}_{0}(|\vec{M}|) + \frac{\mathcal{K}}{M^{2}}
\,(\vec{M}\cdot\vec{e}_{n})^{2} + ... \quad.
\label{phenom_e_int}
\end{equation}
$\mathcal{E}_{0}(|\vec{M}|)$ contains the isotropic energy contributions and $\vec{e}_{n}$ 
represents the direction of the easy-axis. The anisotropy field is then given by 
the first order derivative of $\mathcal{E}_\text{Band}$ with respect to $\vec{M}$ 
(the longitudinal component does not affect the dynamics within the LLG): 
\begin{equation}
\begin{split}
\vec{B}^\text{a} &= -\frac{\partial\mathcal{E}_\text{Band}}{\partial\vec{M}}\quad,\\
                 &= -\frac{2\mathcal{K}}{M^{2}}\,({\vec{M}\cdot\vec{e}_{n}})
                 \,\vec{e}_{n}\quad.  
\end{split}
\label{anisotropy_field}
\end{equation}
Second, the Gilbert damping $(\underline{\mathcal{G}})$ and nutation $(\underline{\mathcal{I}})$ 
tensors shown in section~\ref{LLG_section} are rank-2 tensors, which can be split into 
a symmetric part (labeled with the superscript ${s}$) and an anti-symmetric part (labeled 
with the superscript ${a}$). Moreover, due to the uniaxial symmetry, the Gilbert damping 
tensor has the following structure:
\begin{equation}
\underline{\mathcal{G}} = -\frac{1}{\gamma M}
\left( 
\begin{array}{ccc}
\mathcal{G}^{s}_{\parallel}      & -\mathcal{G}^{a}_{\parallel}  &  \mathcal{G}^{a}_{\perp} \\
\mathcal{G}^{a}_{\parallel}      &  \mathcal{G}^{s}_{\parallel}  & -\mathcal{G}^{a}_{\perp} \\
-\mathcal{G}^{a}_{\perp}         &  \mathcal{G}^{a}_{\perp}      &  \mathcal{G}^{s}_{\perp}
\end{array} 
\right)\quad.
\end{equation}
The symbol $\parallel$ denotes the spin dynamics parameters describing the transverse components of the 
precessional motion when the spin moment is along the [111] direction in its ground state. 
As the system has uniaxial symmetry, the spin dynamics can be anisotropic, and we introduce 
the symbol $\perp$ to account for this possibility. The nutation tensor has the same structure:
\begin{equation}
\underline{\mathcal{I}} = -\frac{1}{\gamma M}
\left( 
\begin{array}{ccc}
\mathcal{I}^{s}_{\parallel}  & -\mathcal{I}^{a}_{\parallel}      &  \mathcal{I}^{a}_{\perp} \\
\mathcal{I}^{a}_{\parallel}      &  \mathcal{I}^{s}_{\parallel}  & -\mathcal{I}^{a}_{\perp} \\
-\mathcal{I}^{a}_{\perp} &  \mathcal{I}^{a}_{\perp}  &  \mathcal{I}^{s}_{\perp}
\end{array} 
\right)\quad.
\end{equation}
The previous decomposition of Gilbert damping and nutation tensors is identical to the 
one performed on magnetic exchange interactions~\cite{Udvardi:2003,Ebert:2009}. The trace
of the the damping tensor coincides with the conventional Gilbert damping constant for a cubic system~\cite{Gilbert:2004}, 
while the off-diagonal components account for the renormalization of $\gamma$, which controls the precession 
rate. Considering the previous forms for the Gilbert damping and nutation combined with Eqs.~\eqref{LLG_generalized_1_adatom} 
and ~\eqref{chi_pm_susc}, the spin-flip dynamical magnetic susceptibility obtained from the LLG equation reads then: 
\begin{equation}
\chi^\text{LLG}_{+-}(\omega) = \frac{1}{2}\frac{M\gamma}{-\frac{2\mathcal{K}\gamma}{M} -(1+
\mathcal{G}^{a}_{\parallel}+i\mathcal{G}^{s}_{\parallel})\,\omega+(-\mathcal{I}^{s}_{\parallel} + 
i \mathcal{I}^{a}_{\parallel})\,\omega^{2}}\quad.
\label{chi_pm_llg}
\end{equation}
The resonance frequency is defined as $\frac{\partial\text{Im} \chi^\text{LLG}_{+-}(\omega)}
{\partial\omega}\big{|}_{\omega^\text{LLG}_\text{res}} = 0$. In absence of nutation, 
it can be computed analytically and is given by: 
\begin{equation}
\omega^\text{LLG}_\text{res} = -\frac{\gamma}{\sqrt{1 + \big(
\mathcal{G}^{s}_{\parallel}\big)^{2} + 2\mathcal{G}^{a}_{\parallel}+ 
\big(\mathcal{G}^{a}_{\parallel}\big)^{2}}}\frac{2\mathcal{K}_\text{susc}}
{M_\text{s}}\quad.
\end{equation}
The latter can be written in terms of the effective gyromagnetic ratio as: 
\begin{equation}
\omega^\text{LLG}_\text{res} = -\frac{\gamma_\text{eff}}{\sqrt{1+
\left(\frac{\mathcal{G}^{s}_{\parallel}}{1+\mathcal{G}^{a}_{\parallel}}\right)^{2}}}
\frac{2\mathcal{K}_\text{susc}}{M_\text{s}}\quad, \quad \text{with} \quad
\gamma_\text{eff} = \frac{\gamma}{1+\mathcal{G}^{a}_{\parallel}}\quad.
\label{gamma_effective}
\end{equation}

\section{Torque method and linear response theory}
\label{appendix_B}
In this appendix, we consider small deviations of the spin moment from the equilibrium direction
and connect the MAE obtained within the torque method and linear response. This will be done
employing the retarded single-particle Green function (GF), which is defined as the resolvent
of the single-particle Hamiltonian $\boldsymbol{\mathcal{H}}(\vec{r}\,)$,
\begin{equation}
  \big(\varepsilon + \iu 0 - \boldsymbol{\mathcal{H}}(\vec{r}\,)\big)\,\boldsymbol{G}(\vec{r}\,,\vec{r}\,';\varepsilon + \iu 0) = \delta(\vec{r}\,-\vec{r}\,') \quad .
  \label{def_GF}
\end{equation}
To keep the notation as light as possible, we do not introduce the partition of space into cells around each atom,
as is customary in the KKR-GF approach. The expressions can easily be generalized to take that aspect into account.
We shall require the following two basic properties (note that the GF is a spin matrix):
\begin{equation}
  \frac{\partial}{\partial\varepsilon}\,\boldsymbol{G}(\vec{r}\,,\vec{r}\,;\varepsilon + \iu 0) =
  - \!\int\!\dd\vec{r}\,'\;\boldsymbol{G}(\vec{r}\,,\vec{r}\,';\varepsilon + \iu 0)\,\boldsymbol{G}(\vec{r}\,',\vec{r}\,;\varepsilon + \iu 0) \quad,
\end{equation}
\begin{equation}
  \frac{\partial}{\partial X}\,\boldsymbol{G}(\vec{r}\,,\vec{r}\,;\varepsilon + \iu 0) =
  \!\int\!\dd\vec{r}\,'\;\boldsymbol{G}(\vec{r}\,,\vec{r}\,';\varepsilon + \iu 0)\,\frac{\partial\boldsymbol{\mathcal{H}}(\vec{r}\,')}{\partial X}\,\boldsymbol{G}(\vec{r}\,',\vec{r}\,;\varepsilon + \iu 0) \quad,\
  \label{derivative_GF_parameter}
\end{equation}
where $X$ is some parameter or quantity upon which the Hamiltonian depends.
Both relations follow trivially from the defining equation of the GF (Eq.~\eqref{def_GF}).
The electronic density of states is given by
\begin{equation}
  \rho(\varepsilon) = -\frac{1}{\pi}\,\text{Im}\,\text{Tr}_{\sigma}\!\int\!\dd\vec{r}\,\;\boldsymbol{G}(\vec{r}\,,\vec{r}\,;\varepsilon + \iu 0) \quad,
  \label{density_of_states}
\end{equation}
from which the connection between the GF and the band energy of the main text $\mathcal{E}_{\text{band}}$ is established.
The spin magnetization density is given by
\begin{equation}
  \vec{M}(\vec{r}\,) = -\frac{1}{\pi}\,\text{Im}\,\text{Tr}_{\sigma}\!\int^{\varepsilon_\text{F}}_{-\infty}\!\!\! \dd\varepsilon\;\vec{\boldsymbol{\sigma}}\,\boldsymbol{G}(\vec{r}\,,\vec{r}\,;\varepsilon + \iu 0) \quad,
\end{equation}
and we make the assumption that the Hamiltonian depends on the direction of the spin magnetization density 
in a coarse-grained way
\begin{equation}
  \boldsymbol{\mathcal{H}}(\vec{r}\,) = \mathcal{H}_0(\vec{r}\,) + {B}_{\text{xc}}(\vec{r}\,)\,\hat{n}(\theta,\varphi)\cdot\vec{\boldsymbol{\sigma}} \quad .
\end{equation}
$\hat{n}(\theta,\varphi)$ being the direction of the exchange-correlation magnetic field. 
Assuming that the easy axis is along the $z$-direction, a small rotation angle $\theta$ in the 
$xz$-plane of $\hat{n}$ results in a torque $\mathcal{T}_{\theta}$ given in Eq.~\eqref{torque_formula_7}. 
Using the definition of the band energy and the density of states (Eqs.~\eqref{e_band_form} 
and~\eqref{density_of_states}), $\mathcal{T}_{\theta}$ can be expressed in terms of the GF as
\begin{equation}
\begin{split}
\mathcal{T}_{\theta} = -\frac{1}{\pi}\,\text{Im}\,\text{Tr}_{\sigma}\!\int\!\dd\varepsilon\!\int\!\dd\vec{r}
\,(\varepsilon - \varepsilon_\text{F})\,\frac{\partial\boldsymbol{G}(\vec{r}\,,\vec{r}\,;\varepsilon + \iu 0)}
{\partial\theta}\quad,
\label{torque_formula_angle}
\end{split}
\end{equation}
Relying on Eq.~\eqref{derivative_GF_parameter}, the first order derivative of the GF with respect 
to $\theta$ can expressed in term of the derivative of $\boldsymbol{\mathcal{H}}(\vec{r}\,)$ which reads: 
\begin{equation}
\begin{split}
\frac{\partial\boldsymbol{\mathcal{H}}(\vec{r}\,)}{\partial\theta} & = 
{B}_{\text{xc}}(\vec{r}\,)\,\frac{\partial\hat{n}(\theta)}
{\partial\theta}\cdot\vec{\boldsymbol{\sigma}}\quad.\\
& = B_\text{xc}(\vec{r}\,)\left[\cos\theta\,\boldsymbol{\sigma}_{x}-\sin\theta\,\boldsymbol{\sigma}_{z}\right]
\end{split}
\label{Der_single_H}
\end{equation}
The combination of the previous equation with Eq.~\eqref{derivative_GF_parameter} and 
Eq.~\eqref{torque_formula_angle} leads to the following expression for the torque: 
\begin{equation}
\mathcal{T}_{\theta} 
= -\frac{1}{\pi} \,\text{Im}\,\text{Tr}_{\sigma}\int^{\varepsilon_\text{F}}_{-\infty}
\dd\varepsilon\int\dd\vec{r}\,B_\text{xc}(\vec{r}\,)\left[\cos\theta\,\boldsymbol{G}(\vec{r},\vec{r},\varepsilon)\,\boldsymbol{\sigma}_{x}-\sin\theta\,\boldsymbol{G}(\vec{r},\vec{r},\varepsilon)\,\boldsymbol{\sigma}_{z}\right]\quad.
\label{torque_expanded_G1}
\end{equation}
The previous expression was obtained after performing a partial energy integration. 
Furthermore, considering a small rotation angle, then $\boldsymbol{G}(\vec{r},\vec{r},\varepsilon)$, 
{\it i.e.} the Green function for the rotated $\vec{B}_\text{xc}$ is related to the 
unperturbed Green function $\boldsymbol{G}_{0}(\vec{r},\vec{r},\varepsilon)$ (with 
$\vec{B}_\text{xc}(\vec{r}) \parallel z$-axis) via a Dyson equation: 
\begin{equation}
\boldsymbol{G}(\vec{r},\vec{r},\varepsilon) \approx \boldsymbol{G}_{0}(\vec{r},\vec{r},\varepsilon) + \int\!\dd\vec{r}^{\,\prime} \boldsymbol{G}_{0}(\vec{r},\vec{r}^{\,\prime},\varepsilon)\,\Delta\vec{B}_\text{xc}(\vec{r}^{\,\prime})\cdot\vec{\boldsymbol{\sigma}}
\,\boldsymbol{G}_{0}(\vec{r}^{\,\prime},\vec{r},\varepsilon)\quad. 
\label{G_Bx_expansion}
\end{equation}
$\Delta\vec{B}_\text{xc}(\vec{r})$ being the change in the exchange-correlation spin-splitting given by:
\begin{equation}
\begin{split}
\Delta\vec{B}_\text{xc}(\vec{r}\,) & = B_\text{xc}(\vec{r\,})\,\left(\sin\theta,0,\cos\theta-1\right)\quad,\\
&\approx B_\text{xc}(\vec{r}\,)\,\left(\theta,0,-\frac{\theta^{2}}{2}\right)\quad.
\end{split}
\end{equation}
Then, the expression of $\boldsymbol{G}(\vec{r},\vec{r},\varepsilon)$ from Eq.~\eqref{G_Bx_expansion} is 
plugged back into Eq.~\eqref{torque_expanded_G1} and $\cos\theta$ and $\sin\theta$ are 
expanded for small $\theta$ as well (retaining linear terms), resulting in the following
from for the torque:
\begin{equation}
\begin{split}
\mathcal{T}_{\theta} 
& = -\frac{1}{\pi}\,\text{Im}\,\text{Tr}_{\sigma}\int^{\varepsilon_\text{F}}_{-\infty}\!
\dd\varepsilon\,\int\!\dd\vec{r}\,B_\text{xc}(\vec{r}\,)\int\dd\vec{r}^{\,\prime}\left[\boldsymbol{\sigma}_{x}\, 
\boldsymbol{G}_{0}(\vec{r},\vec{r}^{\,\prime},\varepsilon)B_\text{xc}(\vec{r}^{\,\prime})
\,\boldsymbol{\sigma}_{x}\,\boldsymbol{G}_{0}(\vec{r}^{\,\prime},\vec{r},\varepsilon)\right]\theta\\
& +\frac{1}{\pi}\,\text{Im}\,\text{Tr}_{\sigma}\int^{\varepsilon_\text{F}}_{-\infty}\!
\dd\varepsilon\,\int\!\dd\vec{r}\,B_\text{xc}(\vec{r}\,)\boldsymbol{\sigma}_{z}\,
\boldsymbol{G}_{0}(\vec{r},\vec{r},\varepsilon)\,\theta\quad.\\
& = \int\!\dd\vec{r}\,B_\text{xc}(\vec{r}\,)\left[\int\!\dd\vec{r}^{\,\prime}
\chi^\text{KS}_{xx}(\vec{r},\vec{r}^{\,\prime},0)
\,B_\text{xc}(\vec{r}^{\,\prime})-M(\vec{r}\,)\right]\theta \quad.
\end{split}
\label{torque_expanded_G2}
\end{equation}
$\chi^\text{KS}_{xx}(\vec{r},\vec{r}^{\,\prime},0)$ is the static Kohn-Sham magnetic 
susceptibility and $M(\vec{r}\,)$ is the magnetization density. Using the definition 
of the spin-flip Kohn-Sham magnetic susceptibility given in Eq.~\eqref{chi_pm_susc} 
in the static limit ({\it i.e.} $\chi^\text{KS}_{xy}(\vec{r},\vec{r}^{\,\prime},0) = 
\chi^\text{KS}_{yx}(\vec{r},\vec{r}^{\,\prime},0) = 0$) and $x$ and $y$-directions are 
equivalent due to uniaxial symmetry), the torque reads: 
\begin{equation}
\mathcal{T}_{\theta} = \int\!\dd\vec{r}\,B_\text{xc}(\vec{r}\,)
\left[2\chi^\text{KS}_{+-}(\vec{r},\vec{r}^{\,\prime},0)\,B_\text{xc}(\vec{r}^{\,\prime})
-M(\vec{r}\,)\right]\theta \quad.
\label{torque_expanded_G3}
\end{equation}
The spin-splitting and the transversal exchange-correlation kernel
$K_{\perp}^\text{xc}(\vec{r}\,)$ are related via~\cite{Samir:2010,Manuel:2015}:
\begin{equation}
B_\text{xc}(\vec{r}\,) = \frac{K_{\perp}^\text{xc}(\vec{r}\,) M(\vec{r}\,)}{2}\quad.
\end{equation}
To obtain a simple result, we coarse-grain the exact equations by integrating out 
the spatial dependence and work with effective scalar quantities. This allows us
to write the transversal exchange-correlation kernel as:
\begin{equation}
K_{\perp}^\text{xc} = \left(\chi_{+-}^\text{KS}(0)\right)^{-1}-\chi_{+-}^{-1}(0)\quad.
\end{equation}
Plugging the two previous expressions into the coarse-grained form of Eq.~\eqref{torque_expanded_G3}, 
$\mathcal{T}_\theta$ can be written in terms of the static spin-flip magnetic susceptibilities 
(Kohn-Sham and enhanced) as: 
\begin{equation}
\mathcal{T}_{\theta} = -\frac{M^{2}}{2}\left[\chi_{+-}^{-1}(0) - \chi_{+-}^\text{KS}(0)
\,\chi_{+-}^{-2}(0) \right]\,\theta\quad.
\label{torque_theta}
\end{equation}
On one hand, considering that $\chi_{+-}(0)$ (static limit) obtained from TD-DFT relates to 
$\mathcal{K}_\text{Susc}$ via $\chi_{+-}(0) = \frac{M^{2}}{4\mathcal{K}_\text{Susc}}$, 
Eq.~\eqref{torque_theta} can be recast into: 
\begin{equation}
\mathcal{T}_{\theta} = -\left(2\mathcal{K}_\text{Susc} - \frac{8\chi^\text{KS}_{+-}(0) 
\mathcal{K}^{2}_\text{Susc}}{M^{2}}\right)\theta\quad.
\label{torque_final_theta}
\end{equation}
On the other hand, the torque $\mathcal{T}_{\theta}$ is also given by the first order derivative 
of the phenomenological form of the band energy as: 
\begin{equation}
\begin{split}
\mathcal{T}_{\theta} & = \frac{\partial\mathcal{E}_\text{Band}}{\partial\theta}\quad,\\
& = -\,\mathcal{K}_{\text{Torque}}\sin2\theta\quad.
\end{split}
\end{equation}
After expanding for a small angle, $\mathcal{T}_{\theta}$ reads: 
\begin{equation}
\mathcal{T}_{\theta} = -2\,\mathcal{K}_{\text{Torque}}\,\theta\quad.
\label{Torque_ksusc_2}
\end{equation}
The connection between $\mathcal{K}_{\text{Torque}}$ and $\mathcal{K}_\text{Susc}$ 
shown in Eq.~\eqref{renormalized_MAE} of the main text can be established when 
comparing Eq.~\eqref{torque_final_theta} and Eq.~\eqref{Torque_ksusc_2}.

\bibliography{mylib_paper.bib}

\end{document}